\documentclass[pdflatex,sn-basic]{sn-jnl}% Basic Springer Nature Reference Style/Chemistry Reference Style
\usepackage{graphicx}%
\graphicspath{{./Images/}}
\usepackage{multirow}%
\usepackage{amsmath,amssymb,amsfonts}%
\usepackage{amsthm}%
\usepackage{mathrsfs}%
\usepackage[title]{appendix}%
\usepackage{xcolor}%
\usepackage{textcomp}%
\usepackage{manyfoot}%
\usepackage{booktabs}%
\usepackage{algorithm}%
\usepackage{algorithmicx}%
\usepackage{algpseudocode}%
\usepackage{listings}%
\theoremstyle{thmstyleone}%
\theoremstyle{thmstyletwo}%

\theoremstyle{thmstylethree}%

\usepackage{subcaption}

\begin{document}

\title[Article Title]{An Optimal Contact-Mechanically Consistent and Flow-Separation Adapted Modeling of Vocal Fold 
Dynamics}


\author[1,2]{\fnm{Sardar Nafis} \sur{Bin Ali}}\email{binsarda@msu.edu}

\author*[2]{\fnm{Maryam} \sur{Naghibolhosseini}}\email{naghib@msu.edu}

\author[1]{\fnm{Mohsen} \sur{Zayernouri}}\email{zayern@msu.edu}

\affil[1]{\orgdiv{Department of Mechanical Engineering}, \orgname{Michigan State University}, \orgaddress{\street{428 S. Shaw Lane}, \city{East Lansing}, \postcode{48824}, \state{MI}, \country{USA}}}

\affil*[2]{\orgdiv{Department of Communicative Sciences and Disorders}, \orgname{Michigan State University}, \orgaddress{\street{1026 Red Cedar Rd}, \city{East Lansing}, \postcode{48824}, \state{MI}, \country{USA}}}
\abstract{\textbf{Objective:}
Single mass-spring-damper models of vocal folds have been shown to be effective in simulating vocal fold vibrations without added complexity. However, single degree-of-freedom models cannot sustain vocal fold oscillation in the presence of structural damping unless the source-tract interaction is taken into account. Moreover, existing lumped models struggle to accurately simulate the vocal fold closure during phonation. This study aims to develop a reliable and simplified model of phonation with a single degree of freedom to simulate the sustained oscillation in a damped system without needing to incorporate the vocal tract model. Additionally, the proposed model can maintain the vocal fold closure in a manner consistent with the physics of phonation, addressing a longstanding challenge in existing lumped models.\\
\textbf{Methods:}
High-speed videoendoscopy (HSV) data were collected from four normophonic subjects (two male and two female) during the production of sustained vowel /i/. After deriving the model’s governing equations, the glottal area waveform (GAW), extracted using a deep learning-based image segmentation technique, is utilized for finding the optimum model parameters employing the particle swarm optimization algorithm. An additional resistance force is incorporated into the model to compensate for the flow separation, which produces an imbalance of forces, required for the sustained oscillation. Additionally, an external structural force is added during the closure to sustain the closure. The 4th-order Runge-Kutta method is employed to solve the governing equations of the proposed model, ensuring an enhanced stability and accuracy for numerical calculations.\\
\textbf{Results:} The model parameters were successfully optimized for individual subjects using particle swarm optimization, resulting in normalized errors of less than 3\% between the experimental and simulated glottal area waveforms. The proposed model accurately reproduced subject-specific vocal fold vibrations and vocal fold closure in agreement with the experimental data.\\
\textbf{Conclusions:} The proposed model is robust framework for simulating the sustained phonation, without requiring the complex analysis of source-tract coupling. Furthermore, the model achieves this using a minimal set of parameters while capturing the key biomechanical and aerodynamic mechanisms of phonation and reducing computational cost.}

\keywords{Phonation Dynamics, Contact Mechanics, Flow Separation Force, Vocal Fold Closure, Mutli-Physics Lumped Modeling}

\maketitle

\newpage
\pagebreak
\section*{Nomenclature} 
\begin{tabbing}
\hspace{4cm} \= \hspace{2cm} \= \kill
$A_g$ \> = \> glottal area \\
$A_{g0}$ \> = \> glottal area at the equilibrium position of the vocal folds \\
$a$ \> = \> acceleration of the vocal fold \\
$C$ \> = \> damping coefficient of the vocal fold \\
$C_{f}$ \> = \> damping amplification factor of the vocal fold \\
$C_{max}$ \> = \> maximum damping coefficient of the vocal fold \\
$C_{min}$ \> = \> minimum damping coefficient of the vocal fold \\
$d$ \> = \> thickness of the vocal fold \\
$F_a$ \> = \> aerodynamic force \\
$F_d$ \> = \> damping force \\
$F_e$ \> = \> elastic force \\
$F_{f.s.}$ \> = \> flow separation force \\
$F_i$ \> = \> inertial force \\
$F_r$ \> = \> reaction force due to the collision of vocal folds \\
$F_{s.a.}$ \> = \> value of the aerodynamic force corresponding to x=$x_c$\\
$F_{s.c.}$ \> = \> structural contact force \\
$K$ \> = \> elasticity of the vocal fold \\
 $\bar{K}_{jv}$ \> = \> $j^{th}$ slope for velocity update in Runge-Kutta method \\
$\bar{K}_{jx}$ \> = \> $j^{th}$ slope for position update in Runge-Kutta method \\
$l$ \> = \> length of the vocal fold \\
$l_e$ \> = \> effective length of the vocal fold over which the aerodynamic force works \\
$M$ \> = \> mass of the vocal fold \\
$P_g$ \> = \> intraglottal pressure \\
$P_{g1}$ \> = \> pressure at inlet of the glottis \\
$P_{g2}$ \> = \> pressure at outlet of the glottis \\
$P_s$ \> = \> sub-glottal pressure \\
$Q_g$ \> = \> intraglottal volumetric flow rate \\
$S$ \> = \> scaling factor \\
$t$ \> = \> time\\
$t_c$ \> = \> closure duration\\
$v$\> = \> velocity of the vocal fold\\
$v_g$\> = \> velocity of air\\
$v_i$\> = \> velocity of the vocal fold at iteration i\\
$v_{i+1}$\> = \>velocity of the vocal fold at iteration i+1\\
$x$\> = \> displacement of the vocal fold \\
$x_0$\> = \> critical displacement of the vocal fold \\
$x_c$\> = \> subcritical displacement of the vocal fold \\
$x_i$\> = \> displacement of the vocal fold at iteration i\\
$x_{i+1}$\> = \>displacement of the vocal fold at iteration i+1\\
$\mu$ \> = \> dynamic viscosity of air \\
$\rho$ \> = \> density of air \\
\end{tabbing}
\pagebreak

\section{Introduction} \label{introduction}
\quad A functional voice is the most effective way to communicate and express thoughts. Coordinated and unobstructed vibration of vocal folds is essential for producing voice properly. Conversely, conditions such as laryngitis, vocal nodules, and vocal cord paralysis  reduce speech quality and impair communication by affecting vocal fold dynamics. Understanding the mechanisms of healthy voice production and how it differs in disorders is an important aspect of voice research and its clinical applications. Numerical modeling of vocal fold dynamics is an efficient and convenient method for investigating these phenomena. The working principle of these models is based on the classical myoelastic–aerodynamic theory developed by \citeauthor{cm1} (\citeyear{cm1}), although the theory is simplified and lacks detailed quantitative formulations. This theory explains how the interaction of muscle tension, tissue elasticity, and airflow causes the vocal folds to vibrate, which produces voice.\\

Vocal fold vibration is a complex process influenced by various factors, including tissue dynamics, aerodynamic forces, and muscular activities (\citeauthor{i1} \citeyear{i1}). However, numerical models can depict the vibrational characteristics of the vocal folds with high precision while containing fewer details than the actual physical structure. Modeling the vibration of vocal folds involves representing it as a structural system affected by both external and internal forces. To study these dynamics, researchers have developed models at different levels of complexity. The finite element method (FEM) treats the vocal folds as a continuum and solves for stress–strain and deformation fields within a defined geometry representing the vibrating tissues. Such models can be constructed in either simplified two-dimensional forms  (\citeauthor{i2} \citeyear{i2}; \citeauthor{i4} \citeyear{i4}; \citeauthor{i3} \citeyear{i3}) or more realistic three-dimensional configurations (\citeauthor{i5} \citeyear{i5}; \citeauthor{i8} \citeyear{i8}; \citeauthor{i6} \citeyear{i6}; \citeauthor{i7} \citeyear{i7}). Although FEM models provide a more accurate representation of the structural properties of the vocal folds, they are computationally expensive due to the need for fine mesh discretization and the solution of large sets of coupled equations. A simpler alternative is the lumped-element model, where the system’s mass is represented as either concentrated at a single point or distributed across multiple points. The viscoelastic properties of such systems are modeled with various arrangements of springs and dampers. The simplest structural lumped-element model of the vocal fold can be characterized by just one mass, allowing only a single degree of freedom. However, the structure of such systems cannot inherently produce asymmetry within the glottal geometry. The asymmetry required for sustained oscillation in these models can arise from the interaction between the vibration source and the vocal tract through acoustic loading (\citeauthor{i11} \citeyear{i11}). This coupling introduces a phase delay between vocal fold velocity and the aerodynamic force, generating the asymmetry necessary for self-sustained vibration. Following this methodology, \citeauthor{i10} (\citeyear{i10}),  \citeauthor{c9} (\citeyear{c9}), \citeauthor{i9} (\citeyear{i9}), \citeauthor{c6} (\citeyear{c6}),  formulated their single mass model. Although these models provide an accurate description of vocal fold vibration, incorporating source–tract coupling makes the frameworks quite complex. Sustained oscillation in single mass models can also be achieved through other types of force asymmetry. \citeauthor{i12} (\citeyear{i12}) achieved sustained oscillation in their single-mass model by introducing a negative damping force to offset the energy lost through damping. Although they explained the similarity between adding a negative damping force and the action of aerodynamic forces in higher–degree-of-freedom models, the introduction of negative damping contradicts the principles of dynamics.  \citeauthor{i13} (\citeyear{i13}) showed that sustained oscillation in a one-mass vocal fold model can be achieved by modeling flow separation and suction forces using the free streamline theory. This assumes the glottal jet is inviscid and irrotational. The model captures inertial effects at high Reynolds numbers but neglects viscous dissipation and shear-layer dynamics. Its analytical expressions are also complex and rely on idealized assumptions, limiting practical use. The model developed by \citeauthor{e1} (\citeyear{e1}) is able to describe the behavior of the vertical phase difference despite being characterized by a single mass. In this framework, the motion of the lower portion of the vocal fold is modeled explicitly, while the motion of the upper portion is determined by applying a scaling factor and a time delay relative to the lower portion’s movement. However, the calculations of the delay and the scaling factor are obtained through complex frequency-domain analyses. 

Introducing more masses or additional degrees of freedom to a single-mass model can overcome its inability to capture the variation of shape in the glottis.  \citeauthor{c2} (\citeyear{c2}), \citeauthor{i17} (\citeyear{i17}), \citeauthor{e4} (\citeyear{e4}), \citeauthor{i14} (\citeyear{i14}), \citeauthor{i16} (\citeyear{i16}), \citeauthor{e3} (\citeyear{e3}), \citeauthor{e2} (\citeyear{e2}), \citeauthor{i15} (\citeyear{i15})  developed this sort of model with a two-mass system. \citeauthor{c4} (\citeyear{c4}) proposed a model that utilized two mechanical resonators. One resonator models the movement of the vocal fold body, driven by the spatially averaged pressure in the glottal region, while the other models the mucosa, which undergoes rotational motion influenced by the axial pressure gradient. \citeauthor{c5} (\citeyear{c5}) proposed a dual-layer model consisting of three masses for the mechanical modeling of vocal fold aiming to provide a more precise model of vocal fold vibrations than the models with one or two masses. Their model considers separate movements of the body and cover of the vocal folds. \citeauthor{i18} (\citeyear{i18}), \citeauthor{i19} (\citeyear{i19})  implemented a triangular body-cover model that extends the classical body-cover model (\citeauthor{c5} \citeyear{c5} ) by incorporating a triangular medial surface geometry and intrinsic muscle activation, enabling more physiologically realistic simulations of vocal fold vibration.  In their study, \citeauthor{c3} (\citeyear{c3}) featured a three-mass system where they incorporated contact forces to simulate collisions during the closure of vocal folds. Other models incorporate even more masses to get a higher level of detail, such as six-mass system introduced by \citeauthor{i20} (\citeyear{i20}),the ten-mass model proposed by \citeauthor{e5} (\citeyear{e5}), and  twenty-five mass system developed by \citeauthor{e6} (\citeyear{e6}). As additional components and details are incorporated into the model, its accuracy in representing vocal fold vibration improves, but the complexity and the number of parameters and equations to be solved also increase. Nevertheless, the increase in computational and practical complexity often outweighs the precision gained from these additions.\\

In addition to the structural aspects, accurate modeling of the forces involved in phonation ensures the sustained oscillations of the system and captures the correct dynamics of glottal vibration. Aerodynamic forces are essential for driving vocal fold oscillation. These forces can be modeled either through the Navier–Stokes equations (\citeauthor{i21} \citeyear{i21}; \citeauthor{i23} \citeyear{i23}; \citeauthor{i22} \citeyear{i22}) or by simplified Bernoulli-based formulations with viscous losses and pressure recovery (\citeauthor{c2} \citeyear{c2}; \citeauthor{c9} \citeyear{c9}; \citeauthor{c5} \citeyear{c5}). Solving the Navier–Stokes equations is computationally intensive, as it requires discretization over many grid points to capture detailed flow fields. In contrast, Bernoulli-based solvers rely on analytical expressions that are easier to compute but do not resolve the full flow field. While Navier–Stokes models provide higher accuracy, Bernoulli-based approaches are often sufficient to predict vocal fold displacements and capture the qualitative characteristics of vibration (\citeauthor{i24} \citeyear{i24}). However, aerodynamic forces alone are sometimes insufficient to ensure sustained oscillation of the system. Sustained oscillation in a damped structural system is achieved by introducing force asymmetry in each cycle of vibration ( \citeauthor{c1} \citeyear{c1}). In multi-degree-of-freedom models, force asymmetry naturally comes from the aerodynamic force equations through flow separation during the closing phase, caused by the divergent glottal shape generated by the mucosal wave produced by the vertical phase difference (\citeauthor{c15} \citeyear{c15}; \citeauthor{c16} \citeyear{c16}). On the other hand, single-degree-of-freedom models must rely on alternative mechanisms of force asymmetry, due to their inherent geometric limitations in capturing the glottal shape variations necessary for flow separation. A simpler and more physically driven solution for these models is the use of a separate flow-separation mechanism that activates during the closing phase of vibration. Another important aspect of vocal fold dynamics is the closed phase, during which the vocal folds collide with each other. 
\citeauthor{i27} (\citeyear{i27}) numerically investigated contact pressure and the patterns of vocal fold contact during phonation and suggested the sensor dimensions required to accurately measure this pressure.
The collision stress for a parabolic vocal fold shape was accurately modeled by \citeauthor{i25} (\citeyear{i25})  using the Hertz model of impact forces. \citeauthor{i26} (\citeyear{i26}) numerically investigated vocal fold contact pressure during phonation and found it to be consistent with Hertzian contact theory. However, this formulation cannot be applied to single–degree-of-freedom models, as they are limited to representing the glottis section as a straight channel. Nevertheless, a different suitable collision model, consistent with physical principles, can be incorporated to represent the closed phase in these models. During this phase, the vocal folds exhibit minimal movement and reduced oscillatory behavior compared to other phases. Accordingly, we propose that the forces incorporated in these models during the closed phase include (1) cancellation of elastic forces, analogous to muscular activation counteracting elasticity, and (2) reaction forces from vocal fold contact, using the conservation of linear momentum, resist further movement of the folds. Incorporating this separate mechanism into the closed phase can also simulate different closure durations.\\

An efficient model is both precise and straightforward, thus simplifying the study, and application of voice science while still relying on accurate theories of physical phenomena. Overly complicated models create difficulties for people to gain insight into the process which makes it challenging for applications. Simple systems require fewer parameters to represent the whole system compared to the complex ones, making optimization easier and reducing computational costs. However, simplification often reduces accuracy. Nevertheless, satisfactory results with simpler models can be obtained by allowing dynamic variations in the model parameters (\citeauthor{c7} \citeyear{c7}). The accuracy of a model must be verified to ensure its validity and to confirm that the correct theories and formulas are applied to accurately replicate vocal fold vibration. A well-balanced model slightly sacrifices accuracy for a significant reduction in complexity.\\

Motivated by the goal of simplifying the understanding of vocal fold dynamics, this study presents the simplest lumped-element model represented by a single mass–spring–damper system. The damping coefficient is presented as a dynamic parameter to account for the nonlinear damping behavior of vocal fold tissues. The vibration is induced by aerodynamic forces,calculated with a simple one-dimensional Bernoulli-based equation that includes viscous losses and pressure recovery. To ensure sustained oscillation of the single–degree-of-freedom model, a flow-separation model is employed. During closure, a structural contact force combining the collision force and the cancellation of elastic forces is utilized.  The oscillatory movements of the vocal folds are calculated using the $4^{\text{th}}$-order Runge-Kutta method, ensuring accuracy and numerical stability. The model parameters are optimized to simulate oscillatory behaviors of the vocal folds for different normophonic subjects during the production of a sustained vowel. Laryngeal high-speed videoendoscopy (HSV) data are used for parameter optimization via implementation of a particle swarm optimization (PSO) algorithm, in addition to evaluating the model’s performance.

\section{Methods} \label{methods}
The data collection protocol and the overview of the data analysis steps are discussed in Section \ref{data}. The construction of the proposed model and the various forces involved in model dynamics are described in Section \ref{modeling}. Section \ref{solver} presents the numerical method used to solve the model’s derived governing differential equations and covers the optimization procedure. Finally, Section \ref{parameters} elaborates on the selection of the model parameters and its evaluation.

 \subsection {Data Collection and Analysis:} \label{data}
 \underline{\textbf{Data Collection:}}  
 
 HSV data and audio signals were collected simultaneously from two male and two female normophonic subjects (see Table \ref{T1}) while producing the sustained /i/ vowel. Audio signals were obtained using an AKG C420 headset condenser microphone connected to a Scarlett 2i2 audio interface (Focusrite, High Wycombe, England). HSV recordings were captured at 4,000 frames per second (fps) with image resolution of $256\times224$ pixels using a Photron FASTCAM mini AX200 monochrome high-speed camera (Photron Inc., San Diego, CA) paired with a flexible nasolaryngoscope.

\begin{table}[htpb]
\centering
\caption{Demographics of the normophonic participants.}
\begin{tabular}{c c c}  \label{T1}
\\ \hline
Subject & Gender & Age\\ \hline
N1 & Male & 49\\
N2 & Male & 30\\
N3 & Female & 35\\
N4 & Female & 22\\ \hline
\end{tabular}

\end{table}

\underline{\textbf{Data Analysis Overview:}} 
a pre-trained U-Net-based convolutional neural network (\citeauthor{c14}, \citeyear{c14}; \citeauthor{i35}, \citeyear{i35}; \citeauthor{i34}, \citeyear{i34}) is used for detecting the glottal area in the obtained HSV data. This network generates a binary mask for each HSV frame with a value of 1 for pixels within the glottal area and 0 for all other regions. The glottal area for each frame is calculated by counting the number of non-zero pixels. The sequence of frames is then analyzed to extract the experimental glottal area waveform (GAW). Subsequently, a model is developed to reproduce GAW patterns consistent with experimental data from subjects. The process begins with constructing a lumped model of vocal folds vibration, followed by computing vocal fold displacements and glottal area change to calculate the GAW during vocal folds’ oscillations. To account for inter-subject variability, model parameters are optimized based on each subject’s experimental data. Finally, the simulated GAW will be compared with experimental data for each subject, as presented in the Section \ref{results}.

\subsection {Modeling framework:} \label{modeling}
The function of the proposed model is based on the interactions between the structural forces exerted by the vocal folds and aerodynamic forces due to the airflow, which is in line with the myoelastic aerodynamic theory of phonation (\cite{cm1}). This study assumes that the vocal folds are symmetrical, possessing identical properties and exhibiting similar dynamic behaviors as this work focuses on modeling normophonic voices. This approach leads to identical forces applied by/on each vocal fold, resulting in similar acceleration, velocity, and displacement of each.

\begin{figure}[htpb]
\begin{center}
\includegraphics[width=1\textwidth]{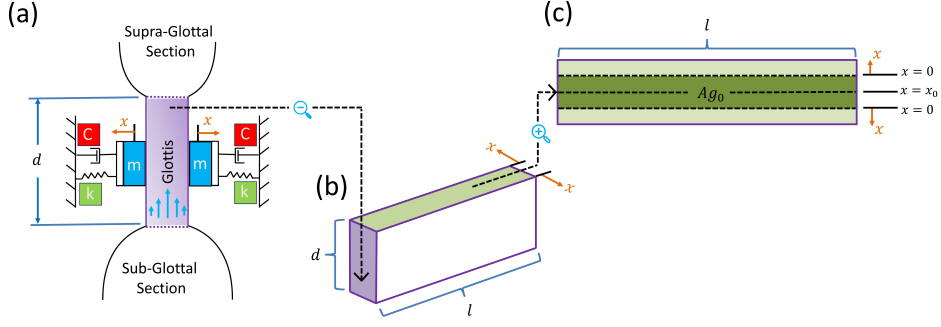}
\caption{Schematic diagram of the lumped model of vocal folds (panel a), 3D view of the glottis geometry (panel b), and superior view of the glottis (panel c). Each vocal fold is simulated with a single mass ($M$), spring ($K$), and damper ($C$) system. $l$ indicates the length and $d$ the thickness of each vocal fold; $x$ represents the displacement from the midline $x_0$ and $A_{g0}$ is the glottal area at the resting position of the vocal folds (at $x = 0$).}
\label{f1}
\end{center}
\end{figure}

The schematic of the proposed model is shown in Fig. \ref{f1}. As can be seen, each vocal fold is simulated using a single mass-spring-damper system (Fig. \ref{f1}a). Accordingly, the internal structural forces within the vocal folds are determined by physical properties of mass ($M$) and viscoelasticity, governed by the spring’s elasticity $K$ and damper’s damping coefficient $C$. The elastic force ($F_e$) depends on the displacement of the vocal fold from its equilibrium position according to:

\begin{equation} \label{eq1}
F_e=-Kx
\end{equation}
In the above equation, the distance of the vocal fold from its equilibrium point is represented by $x$. When the vocal folds move away from each other, the displacement increases in the positive direction, whereas movement in the opposite direction is considered negative. \\
The damping force ($F_d$) is proportional to the velocity of the vocal fold ($v $) as follows:
\begin{equation} \label{eq2}
F_d=-Cv=-C\frac{dx}{dt}
\end{equation}

The oscillation of the vocal folds is powered by the aerodynamic force ($F_a$), originating from the lungs. At the onset of the vibration cycle, it is assumed that the vocal folds are adducted. The buildup of lungs pressure in the subglottal area, along with the elastic force, causes the vocal folds to move apart. Subsequently, the airflow travels through the glottis. As the distance between the folds increases, the available cross-sectional area for airflow also increases. Assuming an inviscid, incompressible, and steady flow, according to Bernoulli's principle, as the flow velocity decreases, the intraglottal  pressure increases, thereby increasing the aerodynamic force that keeps the folds apart. The elastic force then brings the vocal folds back toward the midline, leading to an increase in airflow velocity and a drop in intraglottal pressure, which further helps with the vocal fold closure. As the vocal folds contact, they experience a structural contact force ($F_{s.c.}$) that keeps them closed for a specific duration. When $F_{s.c.}$ is deactivated, the vocal folds can reopen again due to the elastic force and the static subglottal pressure. The vocal folds’ continuous opening and closing in a cyclic manner leads  to the generation of an acoustic signal. A flow separation force ($F_{f.s.}$) is included in the model in order to maintain the sustained oscillation. In the absence of this force, the amplitude of the oscillation will be reduced due to the damping in the system. Thus, the total force acting on each vocal fold, to ensure a sustained oscillation, is expressed as:

\begin{equation} \label{eq3}
Ma=F_{e}+F_{d}+F_{a}+F_{f.s.}+F_{s.c.},
\end{equation}
with $a$ representing the acceleration of the vocal fold.

The magnitude and involvement of these forces in vocal folds’ vibration depend on the displacement of the vocal folds and the areas formed between them. The glottal area ($A_g$) changes as a function of x is formulated as:

\begin{equation} \label{eq4}
A_g=A_{g0}+2lx,
\end{equation}
in which, $A_{g0}$ represents the glottal area when the vocal folds are at their equilibrium position $x=0$ (see Fig. \ref{f1}-c). Considering that the displacement of each vocal fold by $x$ increases the area by $lx$ with respect to $A_{g0}$, the total increase in the area is $2lx$ as captured in Eq. \ref{eq4}.\\

We define the critical displacement $x_0$ as the displacement of the vocal folds beyond their equilibrium position at a point of vocal folds’ full contact, which  makes the glottal area equal to zero. The value of $x_0$ can be calculated by plugging in $x_0$ into Eq. \ref{eq4} as:
\begin{equation} \label{eq5}
\begin{split}
A_g & = A_{g0}+2lx_0=0 \\
\text{hence, } x_0 & = \frac{-A_{g0}}{2l}
\end{split}
\end{equation}

The aerodynamic pressure, generated due to the glottal airflow, acts on the area $ld$ of each vocal fold.  The  aerodynamic force $F_a$ originates from the pressure distribution on the vocal folds, which is assumed to vary linearly, changing from $P_{g1}$ at the inlet to $P_{g2}$  at the outlet (\cite{c9}). As a result, $F_a$ is determined by multiplying the average of $P_{g1}$ and $P_{g2}$ by the area $ld$, over which this force acts upon: 

\begin{equation}  \label{eq6}
F_a =
\begin{cases} 
\frac{(P_{g1}+P_{g2})ld}{2}, & \text{if } x > x_0\\
0, & \text{otherwise}.
\end{cases}
\end{equation}
In Eq. {\ref{eq6}}, $F_a$ vanishes when the displacement is less than or equal to $x_0$, due to the absence of airflow.\\
The magnitudes of $P_{g1}$ and $P_{g2}$ are functions of the intraglottal volumetric flow rate  ($Q_g$). $Q_g$ depends on the sub-glottal pressure ($P_s$) and the airflow resistance through the glottis. This resistance arises from both frictional and turbulence losses. $Q_g$ is calculated based on the assumption that the airflow is inviscid, incompressible, and steady according to the following formula  (\cite{c10}; \cite{c8}; \cite{c9}; \cite{c11}; \cite{c12}):

\begin{equation} \label{eq7}
\frac{0.875 \rho Q_g^2}{2A_g^2}+\frac{12 \mu d l^2 Q_g}{A_g^3}-P_s=0,
\end{equation}

where $\rho$ corresponds to the air density and $\mu$ to the dynamic viscosity. Let us define the following in Eq. \ref{eq7}:
\begin{equation} 
\begin{split}
a & = \frac{0.875 \rho}{2A_g^2} \nonumber \\
b &= \frac{12 \mu d l^2}{A_g^3} \nonumber \\
c &=-P_s \nonumber
\end{split}
\end{equation}

Therefore, the flow rate $Q_g$ can be formulated as:
\begin{equation}  \label{eq8}
Q_g =
\begin{cases} 
\frac{-b+\sqrt{b^2 - 4ac}}{2a}, & \text{if } x > x_0\\
0, & \text{otherwise}.
\end{cases}
\end{equation}

If the displacement of the vocal folds becomes less than $x_0$, then $Q_g=0$ due to the absence of airflow through the glottis.\\

The intraglottal pressure, denoted as $P_g$, emerges as a dynamic pressure due to the airflow velocity change through the glottis. If the flow velocity of the incompressible air through the glottis is expressed as $v_g  = \frac{Q_g}{A_g}$, $P_g$ can be formulated as:
\begin{equation}  \label{eq9}
P_g =
\begin{cases} 
\frac{1}{2} \frac{\rho Q_g^2}{A_g^2}, & \text{if } x > x_0\\
0, & \text{otherwise}.
\end{cases}
\end{equation}

The formulas for the pressures at the inlet ($P_{g1}$) and outlet ($P_{g2}$) of the glottis can be formulated as follows (\cite{c9}; \cite{c10}; \cite{c12})  :
\begin{equation} \label{eq10}
P_{g1}=P_s-1.37P_g
\end{equation}
\begin{equation} \label{eq11}
P_{g2}=-\frac{1}{2}P_g
\end{equation}

$F_a$ varies depending on the vocal folds vibratory phase. If the vocal folds are closed, $F_a$ becomes zero (as shown in Eq. \ref{eq6}). When the vocal folds are fully open, $F_a$ is negative. Conversely, when the vocal folds are adducting and become very close to each other, $F_a$ becomes positive. When the vocal folds are between these two extremes, the aerodynamic force falls between the maximum positive and negative values. The sub-critical distance $x_c$ is defined as the displacement corresponding to an aerodynamic force within this range.\\

\begin{figure}[htbp]
\begin{center}
\includegraphics[width=1\textwidth]{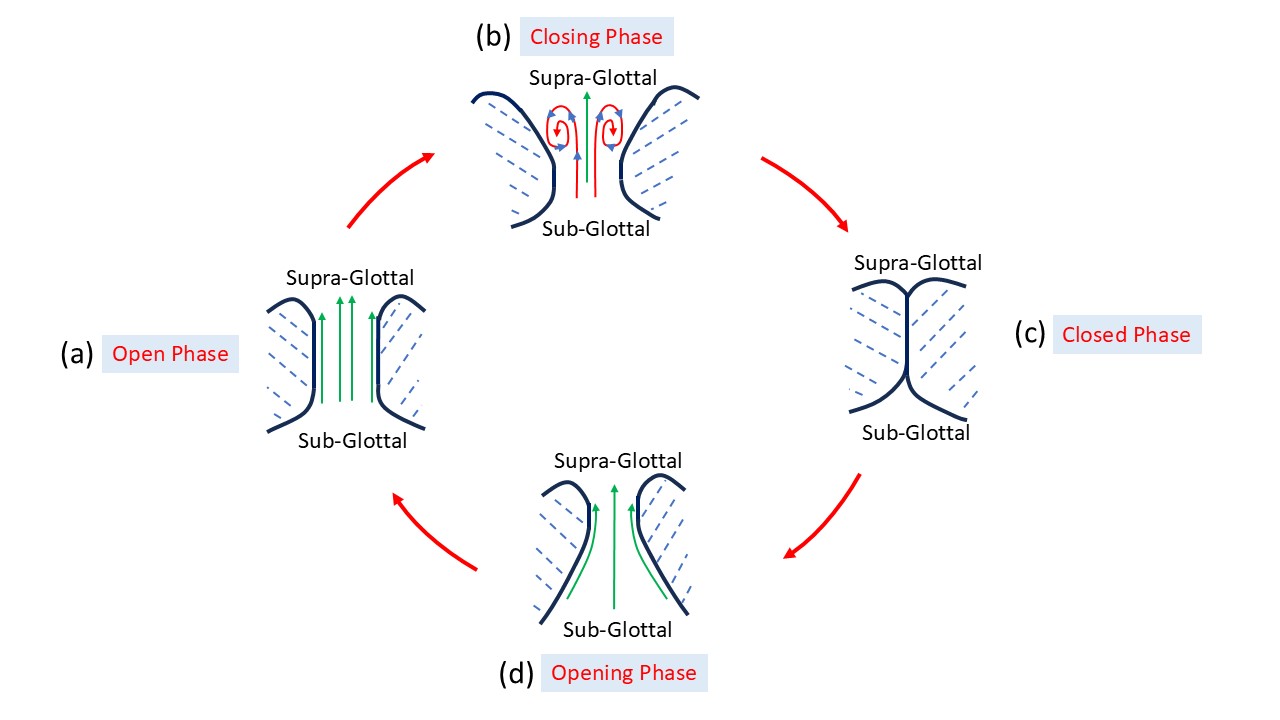}
\caption{Schematic representation of  various glottal geometry formed by the vocal folds during different phases of the glottal cycle: (a) a straight channel-like shape during the open phase, (b) a divergent nozzle-like shape during the closing phase (flow separation occurs in such a configuration), (c) fully adducted vocal folds preventing airflow through $A_g$ throughout the closed phase and (d) a convergent nozzle-like shape during the opening phase.}
\label{f2}
\end{center}
\end{figure}

Each glottal vibratory cycle in the current model is divided into four distinct phases (the opening, open, closing and the closed phase) depending on $x_c$ and $x_0$ and the direction of intraglottal velocity. Fig. \ref{f2} shows the different phases of the glottal cycle. As can be seen, there is a phase difference between the upper and lower edges of the vocal folds during the opening and closing phases (Fig. \ref{f2}b and \ref{f2}d) leading to the generation of mucosal wave (\cite{c15}; \cite{c16}). During the opening phase, the lower edges of the vocal folds abduct first, followed by the subsequent gradual opening of the upper edges with a delay. During the closing phase, the lower edges adduct first and the upper edges follow with a delay. As a result, a converging nozzle-like shape during the opening phase and a diverging nozzle-like shape during the closing phase is created. In multi-degree-of-freedom models, the vocal folds’ sustained oscillation is maintained through a variable glottal geometry. Considering that the current model has one degree of freedom, it produces a constant glottal geometry.   We overcome this constraint by applying the flow separation force $F_{f.s.}$ during the closing phase.\\
The involvement and roles of various forces in the model during different phases of the glottal cycle are described in the following.

\textbf{Open Phase:} During the open phase, the shape of the glottal passage resembles a straight channel as depicted in Fig. \ref{f2}a. During this phase, the displacement of the vocal folds is bigger than subcritical displacement $x_c$ ($x>x_c$; $x_c<0$). If the aerodynamic force at $x_c$ is denoted as $F_{s.a}$, during the open phase, $F_a$ remains less than $F_{s.a.}$ Throughout this phase, the  vocal folds maintain an open configuration with a constant flow area along their length, preventing the formation of any flow separation ($F_{f.s.}=0$). Additionally, $F_{s.c.}$ remains zero since the vocal folds are not in contact.\\

\textbf{Closing Phase:} As $F_a$ increases beyond its value at $x_c$, the closing phase  begins until the vocal folds fully contact at $x=x_0 (x_c \geq x>x_0 )$. In this phase, the vocal folds move toward each other and the distance between them becomes very small but are not in contact yet, hence, $F_{s.c.}=0$. During this phase, the  glottal passage takes the shape of a diverging nozzle (Fig. \ref{f2}b). This glottal geometry causes flow separation due to the adverse pressure gradient in the direction of the flow. The flow separation point gradually moves upstream with time and accordingly, the area impacted by the aerodynamic force decreases over time.  A linear decrease in the length of the vocal folds is incorporated into model to simulate this phenomenon. This linear reduction in the length leads to a proportional decrease in the surface area on which the aerodynamic force acts, as the area is equal to the product of the length and d of the vocal fold. The length over which the aerodynamic force acts is denoted as effective length $l_e$, which we assumed that it follows a linear profile as a function of displacement from $x_c$ in order to model the flow separation during this phase. Therefore: 
\begin{equation} \label{eq12}
l_e=l-\frac{l(x-x_c)}{x_0-x_c}
\end{equation}

The impact of the decreasing effective length can be expressed as a counterbalancing aerodynamic force in the form of $F_{f.s.}$ using the following formula:
\begin{equation}  \label{eq13}
F_{f.s.} =
\begin{cases} 
-\frac{(P_{g1}+P_{g2})}{2} \times \frac{l(x-x_c)}{(x_0-x_c)} \times d, & \text{if } x_c\geq x > x_0 \text{ and } v=\text{``}-\text{ve''} \\
0, & \text{otherwise}.
\end{cases}
\end{equation}

\textbf{Closed Phase:} When the displacement of vocal folds reaches $x_0$, the glottal vibration shifts to the closed phase. During this phase, the vocal folds are in contact with each other $(x \leq x_0 )$, causing $A_g$ to become zero, and hence, the glottal airflow and $F_a$ are zero. At the beginning of the closed phase, the vocal folds move beyond point $x_0$ due to their velocity, leading to their collision and compression. The vocal folds’ velocity $v$ decreases as they compress, and reaches zero at the point of maximum compression, where the damping force $F_d$ becomes zero. Beyond this point and during the closed phase, the structural contact force $F_{s.c.}$ is defined as follows:
\begin{equation}  \label{eq14}
F_{s.c.}= 
\begin{cases} 
F_r-F_e & \text{if } x \leq x_0 \text{ and } v=0\\
0, & \text{otherwise}.
\end{cases}
\end{equation}
The reaction force ($F_r$) in the above formula is due to the collision of the modeled vocal fold with the other fold. Moreover, $F_e$ is subtracted from $F_r$ to overcome the elastic force and prevent the vocal folds from opening too early. Accordingly, the total closure duration of $t_c$ is maintained.\\

\textbf{Discrete formulation of $F_{s.c.}$:} Let the position and velocity of the vocal fold at time $t$ be $x_n$ and $v_n$, respectively. At time $t + dt$, the position and velocity change to $x_{n+1}$ and $v_{n+1}$, respectively. After another time step of $dt$, the position and velocity will update to $x_{n+2}$ and $v_{n+2}$. Using the forward Euler method, the relation between displacements and velocities can be established as follows:
 \begin{equation} \label{eq15}
\begin{split} 
x_{n+1} &=x_n+v_ndt\\
x_{n+2} &=x_{n+1}+v_{n+1}dt
\end{split}
\end{equation}
$F_r$ resulting from the collision of the vocal folds can be considered as the rate of change in the momentum, as follows:
\begin{align}  \label{eq16}
F_{r} &= \frac{M(v_{n+1}-v_{n})}{dt} \notag \\
&= \frac{M(v_{n+1}-2v_n+v_n)}{dt} \notag \\
&= \frac{M(v_{n+1}dt-2v_ndt+v_ndt)}{dt^2} \notag \\
&= \frac{M(v_{n+1}dt-2v_ndt+v_ndt+x_n-x_n)}{dt^2} \notag \\
&= \frac{M(x_n+v_{n}dt+v_{n+1}dt-x_n-2v_ndt)}{dt^2} \\
&= \frac{M(x_{n+1}+v_{n+1}dt-x_n-2v_ndt)}{dt^2} \notag \\
&= \frac{M(x_{n+2}-x_n-2v_ndt)}{dt^2} \notag \\
&= \frac{M(x_{n+2}-x_n)}{dt^2} - \frac{2Mv_n}{dt} \notag \\  \notag
\end{align}

Ultimately, the structural contact force $F_{s.c}$ can be expressed as:
\begin{equation}  \label{eq17}
F_{s.c.}= 
\begin{cases} 
F_r-F_e=  \frac{M(x_{n+2}-x_n)}{dt^2} - \frac{2Mv_n}{dt}+Kx, & \text{if } x \leq x_0 \text{ and } v=0\\
0, & \text{otherwise}.
\end{cases}
\end{equation}

\textbf{Opening Phase:} The vocal folds begin to open following the closed phase $(x_0 < x \leq x_c)$. During this phase, $F_{s.c.}$ is zero and the converging nozzle shape of the glottal geometry prevents flow separation due to a favorable pressure gradient in the direction of the flow, hence, $F_{f.s.} = 0$. Once the opening phase ends $(x > xc)$, the vibratory phase transitions back to the open phase, starting the next cycle of vibration.

\subsection {Numerical Solution and Model Optimization} \label{solver}
The Runge-Kutta $4^{\text{th}}$-order method (RK4) is used for solving the following derived stiff time-dependent differential equations:
\begin{align} 
\frac{dx}{dt} =\dot{x}&=v  \label{eq18} \\
\frac{dv}{dt} = \dot{v}=\frac{\sum F}{M} &=a=\frac{(F_{e}+F_{d}+F_{a}+F_{f.s.}+F_{s.c.})}{M} \label{eq19} 
\end{align}

RK4 is selected due to its higher order of accuracy and stability in numerical calculations in comparison with other competing methods such as the forward and backward Euler methods, Heun’s method, the midpoint method, and the trapezoidal rule. Eq. \ref{eq20} and \ref{eq21} show how $x$ and $v$ are updated at each time step $dt$. In these equations, $x_{i+1}$ and $v_{i+1}$ represent the updated position and velocity, while $x_i$ and $v_i$ represent the position and velocity at the current step. As can be seen, the values of $x$ and $v$ at the new time step are computed based on their values at the previous time step and the weighted average of calculated slopes of $\bar{K}_{jx}$ and $\bar{K}_{jv}$ for $x$ and $v$, respectively. The index $j$ ranges from 1 to 4, corresponding to four intermediate stages (slope evaluations) of the RK4 for $x$ and $v$ variables.

\begin{align}
x_{i+1} &=x_{i}+\frac{1}{6}(\bar{K}_{1x}+2\bar{K}_{2x}+2\bar{K}_{3x}+\bar{K}_{4x}) \label{eq20} \\
v_{i+1} &=v_{i}+\frac{1}{6}(\bar{K}_{1v}+2\bar{K}_{2v}+2\bar{K}_{3v}+\bar{K}_{4v}) \label{eq21}
\end{align}

The formulation of $F_r$ and accordingly $F_{s.c.}$ in Eq. \ref{eq17}  is based on the forward Euler method as it was previously explained. The forward Euler method is a single-step numerical scheme, whereas the RK4 is a multi-step method, offering greater accuracy due to its use of multiple intermediate slope evaluations within each time step. Considering that both the RK4 and forward Euler method are explicit time integration methods, we can consider $x_n$ in Eq. \ref{eq17} as the displacement $x_i$ at the initial step, $x_{n+1}$ as the displacement at an intermediate step, and $x_{n+2}$ as the displacement $x_{i+1}$ at the final step. Accordingly, $F_{s.c.}$ in Eq \ref{eq17} can be formulated as:
\begin{equation}  \label{eq22}
F_{s.c.}= 
\begin{cases} 
F_r-F_e=  \frac{M(x_{i+1}-x_i)}{dt^2} - \frac{2Mv_i}{dt}+Kx , & \text{if } x \leq x_0 \text{ and } v=0\\
0, & \text{otherwise}.
\end{cases}
\end{equation}

The following cost function is defined toward optimizing the model parameters:
\begin{equation} \label{eq23}
\text{Cost Function}= \frac{\sum_{1}^{\text{no. of data points}}(A_{g, \text{exp.}}^2 - A_{g,\text{sim.}}^2) }{\sum_{1}^{\text{no. of data points}}A_{g,\text{exp.}}^2}
\end{equation}

The above cost function calculates the normalized error between the experimental glottal area waveform $A_{g,\text{exp}}$ (estimated from the HSV data) and the simulated glottal area waveform $A_{g,\text{sim}}$ (calculated by the model based on Eq. \ref{eq4}). The objective of the optimization process is to minimize the cost function for different subjects. As the cost function is non-convex, the solution may get stuck at a local minimum instead of converging to the global minimum. To address this issue, particle swarm optimization (PSO) algorithm is employed, as it can efficiently identify the global minimum in a non-convex function (\citeauthor{c13}, \citeyear{c13}).

In the PSO algorithm, multiple particles are used to search for the best solution. Each particle is characterized by a set of parameter values to be optimized and their corresponding cost function value. At the beginning of the optimization process, initial values for all particles and their ranges are defined, and the velocity of all particles is set to zero. PSO refines the search iteratively until it converges toward the optimal solution. At each step, the position of each particle is updated based on its velocity, which is determined by its current velocity vector and the distance vectors from its current location to the local best and global best positions. The local best of each particle refers to its specific set of parameter values that results in its lowest cost function value. On the other hand, the global best solution is determined by selecting the particle having a set of parameter values that yield the lowest cost function value. The cost function values for all particles across all completed iterations are evaluated to determine the global best position. Updating the positions of the particles at each step ensures that the cost function values are reduced compared to the previous step. Once all iterations are completed, the global best position is used for optimal parameters. 

\subsection{Model Parameters}\label{parameters}
\begin{figure}[htpb]
\begin{center}
\includegraphics[width=1\textwidth]{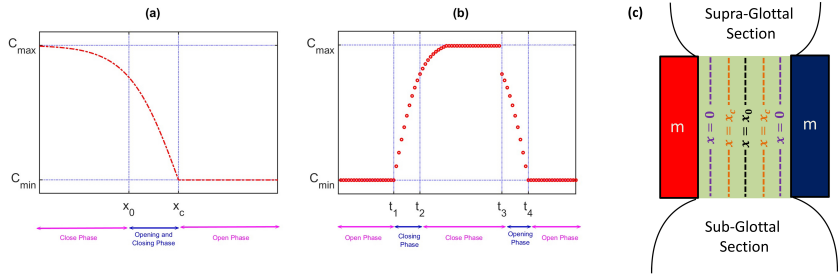}
\caption{Variation of $C$ as a function vocal fold displacement $x$ (panel a) and as a function of time (panel b) across different phases of the glottal cycle. Referring to panel (b), at $t=t_1$, $x=x_c$ (the open phase ends and the closing phase begins); at $t=t_2$, $x=x_0$ (the closing phase ends and the closed phase begins); at $t=t_3$, $x=x_0$ (the closed phase ends and the opening phase begins); at $t=t_4$, $x=x_c$ (the opening phase ends and the open phase begins). Panel (c) illustrates a schematic representation of different positions of the left (colored red) and right (colored blue) vocal folds. }
\label{f3}
\end{center}
\end{figure}

For each subject, all model parameters are kept constant throughout every phase of the vibration cycle except for the damping coefficient $C$. $C$ is simulated nonlinearly to capture the nonlinear damping behavior of vocal fold tissues. The variation of $C$ with respect to the vocal fold displacement is shown in Fig. \ref{f3}. $C$ is varied smoothly throughout each cycle of glottal vibration to reflect changes in the resistance to motion depending on the distance and degree of contact between the vocal folds. The change in $C$ is modeled using the positive portion of the hyperbolic tangent function. This positive segment is scaled by a factor called the damping amplification factor ($C_f$) and shifted by the value of the minimum damping coefficient ($C_{min}$), allowing $C$ to increase smoothly from its minimum value to its maximum ($C_{min}+C_f$). This approach  accounts for vocal folds’ nonlinear damping behavior by utilizing the highly nonlinear characteristics of the hyperbolic tangent function.  Moreover, the finite upper limit of the hyperbolic tangent function bounds the damping coefficient to $C_{max}=C_{min}+C_f$. This constraint prevents $C$ from becoming infinite, which will be physically unrealistic.  Throughout the open phase ($x>x_c, x_c<0$), $C$ remains constant at its lowest value $C_{min}$ due to the low resistance to motion because of a large distance between the vocal folds. During the closing phase ($x_c \geq x>x_0 \text{ and } v=\text{``}-\text{ve''}$), the distance between the vocal folds starts decreasing, and $C$ increases in a hyperbolic tangent manner as a function of the displacement. Throughout the closed phase ($x\leq x_0$ ), $C$ follows the same profile as in the closing phase and reaches its highest value due to the compression of the vocal folds, where the distance of vocal folds from $x_c$ is at its maximum. During the opening phase ($x_0 < x \leq x_c \text{ and } v=\text{``}+\text{ve''}$) the distance between the vocal folds increases, and $C$ follows the same hyperbolic tangent profile but decreases over time. The formula for $C$ can be expressed as follows:
\begin{equation}  \label{eq24}
C =
\begin{cases} 
C_{min}+C_f \times tanh(x_c-x), & \text{if }  x<x_c\\
C_{min}, & \text{otherwise}.
\end{cases}
\end{equation}

An additional parameter, denoted as the scaling factor $S$, is defined to provide consistency between the units of the glottal area waveform from the numerical data (expressed in square millimeters, $\text{mm}^2$) and the experimental data (measured in pixels). $S$ is multiplied with the numerical results for converting the unit of the numerical outcomes into pixels, making the numerical results comparable to the experimental results. Mathematically, it is defined as:
\begin{equation} \label{eq25}
S= \frac{\text{pixels}}{\text{mm}^2}
\end{equation}

\section{Results}\label{results}
Several parameter values used in the model are obtained from existing literature, while others are calculated through the optimization as explained in the Methods section. The values of the parameters used in the simulations are provided in Tables \ref{T2} and \ref{T3}.
\begin{table}[htpb]
\centering
\caption{A list of parameters that remain constant for each subject in the simulations.}
\begin{tabular}{c c c}  \label{T2}
\\ \hline
Parameter & Value & Unit\\ \hline
$d$ & 0.32 & cm\\
$C_{min}$ & 0 & $\frac{\text{gm}}{s}$\\
$\mu$ & $1.8 \times 10^{-4}$ & Poise\\
$\rho$ & $1.3 \times 10^{-3}$ & $\frac{\text{gm}}{\text{cm}^3}$\\
$A_{g0}$ & 0.05 & $\text{cm}^2$\\
$F_{s.a.}$ & 0 & dyne\\ \hline
\end{tabular}

\end{table}

The closure time ($t_c$) for each subject is different, and they are calculated from the experimental glottal area waveforms. Like $t_c$, $l$ also varies depending on the subjects. The length of the vocal fold is considered to be 1.6 cm for male subjects and 1 cm for female subjects based on the study by \citeauthor{e7} (\citeyear{e7}). Moreover, this study indicateed that the density of the vocal folds exhibits minimal variation across subjects. As a result, in this work, the mass per unit length is regarded as constant for all subjects. The mass of the vocal folds is set at 0.12 grams for men and 0.075 grams for women. The ratio of male to female vocal fold mass is 1.6, equal to the ratio of their vocal fold length.

\begin{table}[htpb]
\centering
\caption{A list of different parameters used in the simulations; these parameters remain fixed for each subject but vary between subjects.}
\begin{tabular}{c c c c}  \label{T3}
\\ \hline
Subject & $m$(gm) & $l$(cm) & $t_c(ms)$\\ \hline
N1 & 0.12 & 1.6 & 5.25\\
N2 & 0.12 & 1.6 & 4\\
N3 & 0.075 & 1 & 2.75\\
N4 & 0.075 & 1 & 1.5\\ \hline
\end{tabular}

\end{table}

The experimental glottal area waveform (GAW) for each participant consists of 201 points, selected from the HSV frames for the sustained /i/ vowel phonation, during which the glottal area amplitude, cycle length, and closure time show minimal variations. The time interval between these points is $\frac{1}{4000}$ \textit{s}, corresponding to the inverse of the HSV recording frame rate. In the numerical calculations, solutions are computed at intervals of $\frac{1}{128\times4000}$ \textit{s}—128 times smaller than the time interval of experimental data. This finer resolution is necessary to maintain the stability of the highly stiff system and reduce numerical errors. After numerical calculations, the results are down sampled to match the experimental data- 201 points are selected from the total of 25,601 simulated points by sampling at intervals of $\frac{1}{4000}$ seconds. This ensures that the numerical and experimental data have the same number of points at the same time interval for direct comparison.

\begin{figure}[htbp]
\centering

\begin{subfigure}{0.45\textwidth}
    \centering
    \includegraphics[width=\linewidth]{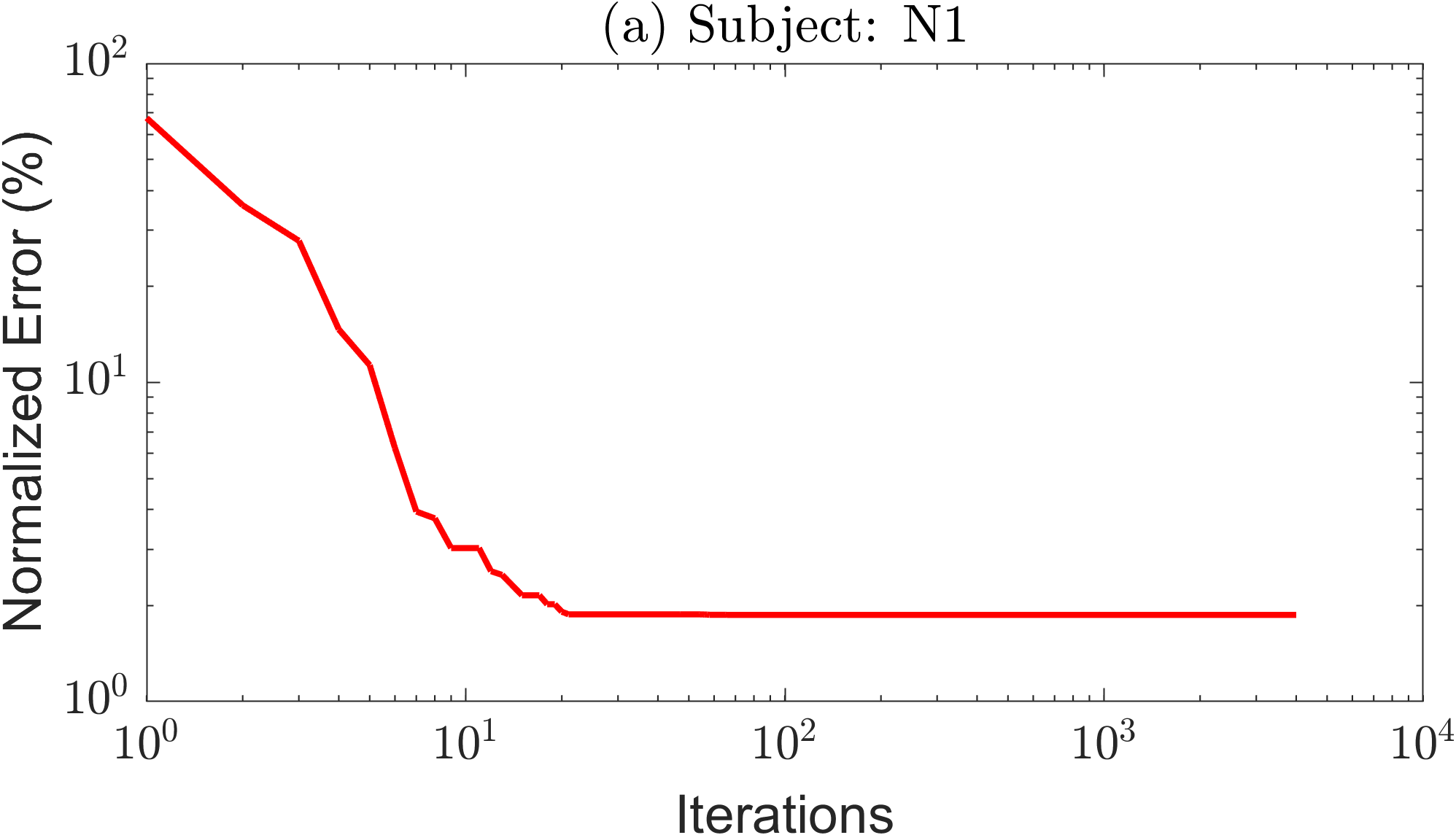}
    \caption{}
\end{subfigure}
\hfill
\begin{subfigure}{0.45\textwidth}
    \centering
    \includegraphics[width=\linewidth]{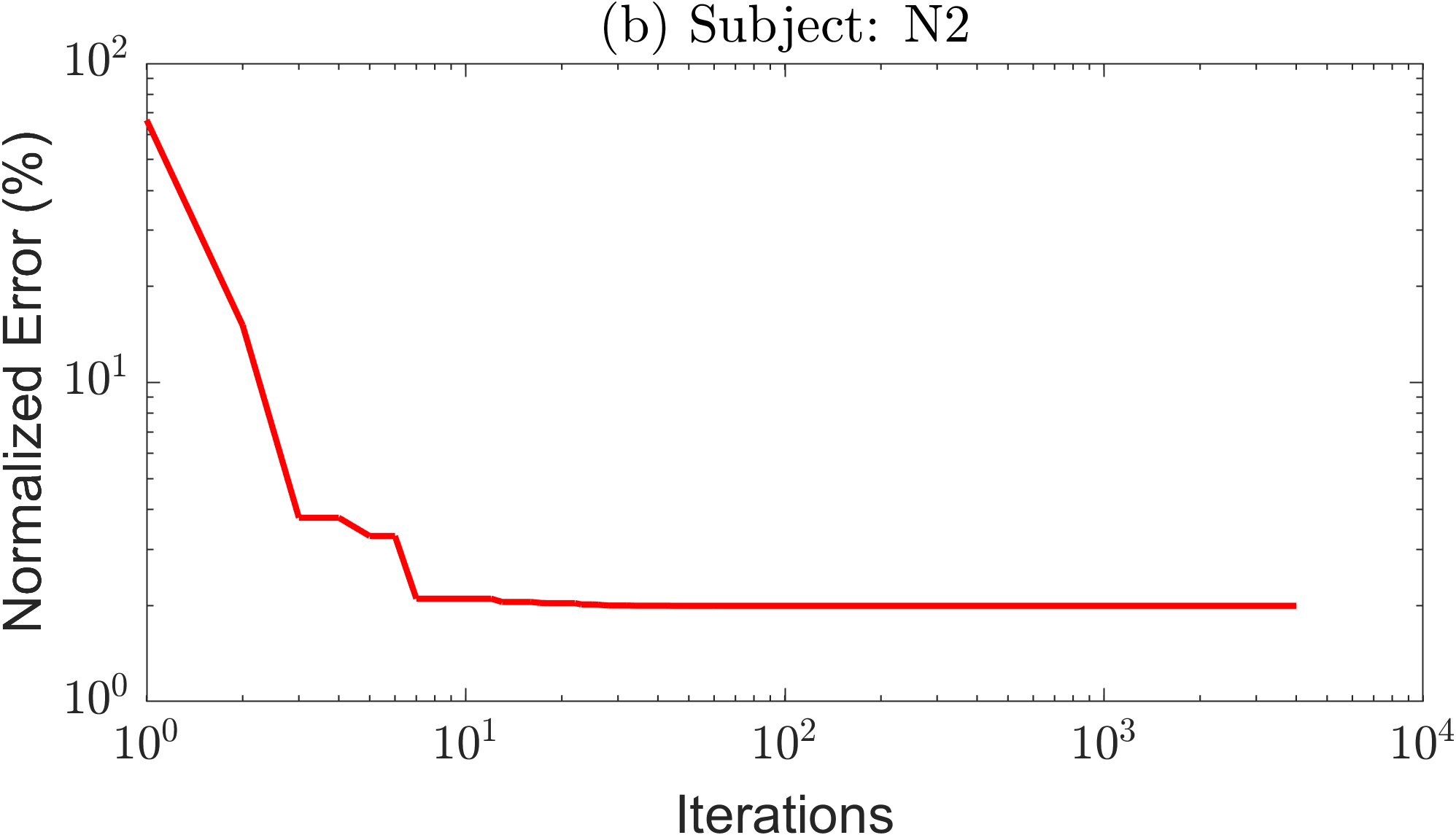}
    \caption{}
\end{subfigure}

\vspace{0.5cm}

\begin{subfigure}{0.45\textwidth}
    \centering
    \includegraphics[width=\linewidth]{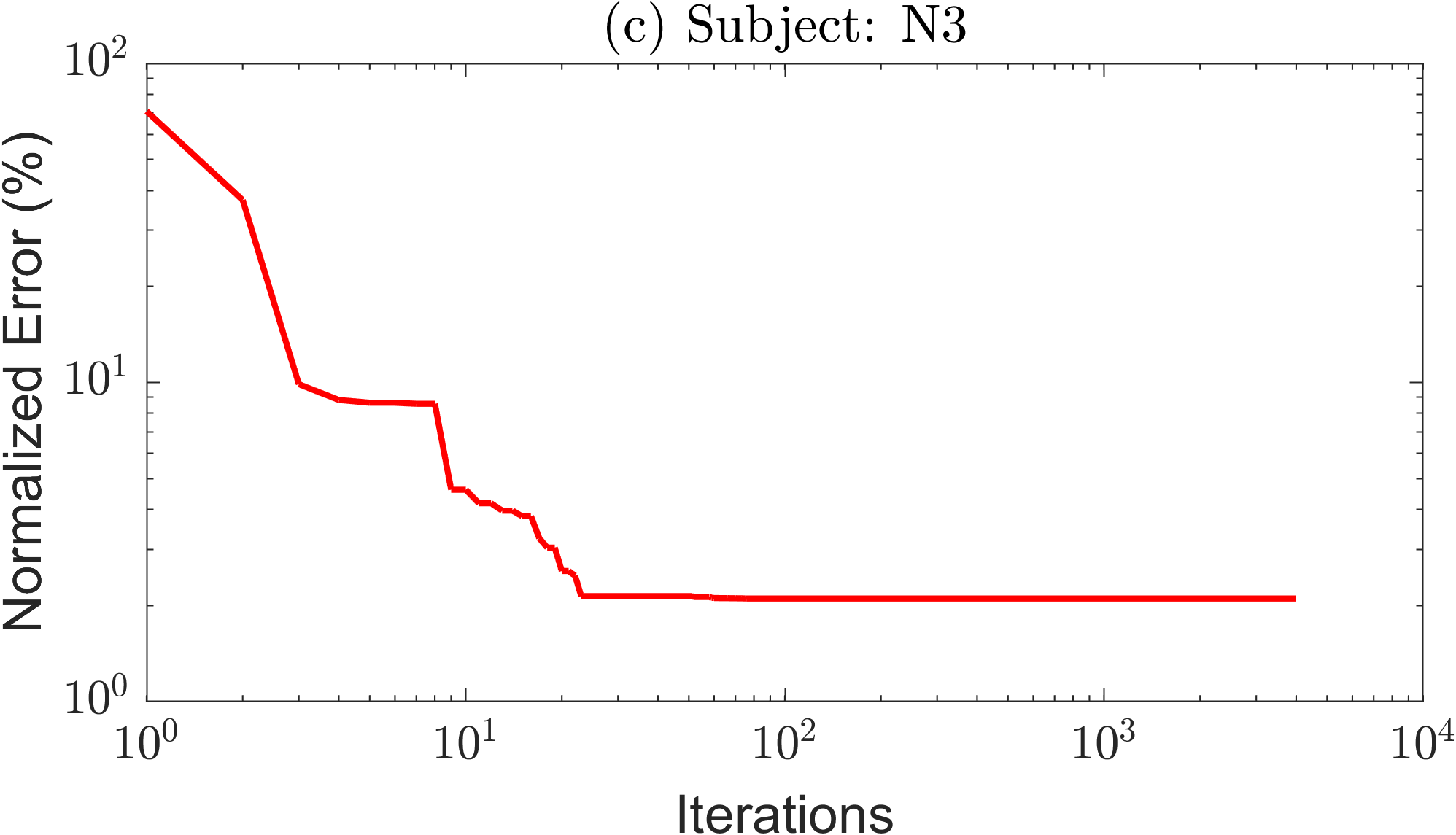}
    \caption{}
\end{subfigure}
\hfill
\begin{subfigure}{0.45\textwidth}
    \centering
    \includegraphics[width=\linewidth]{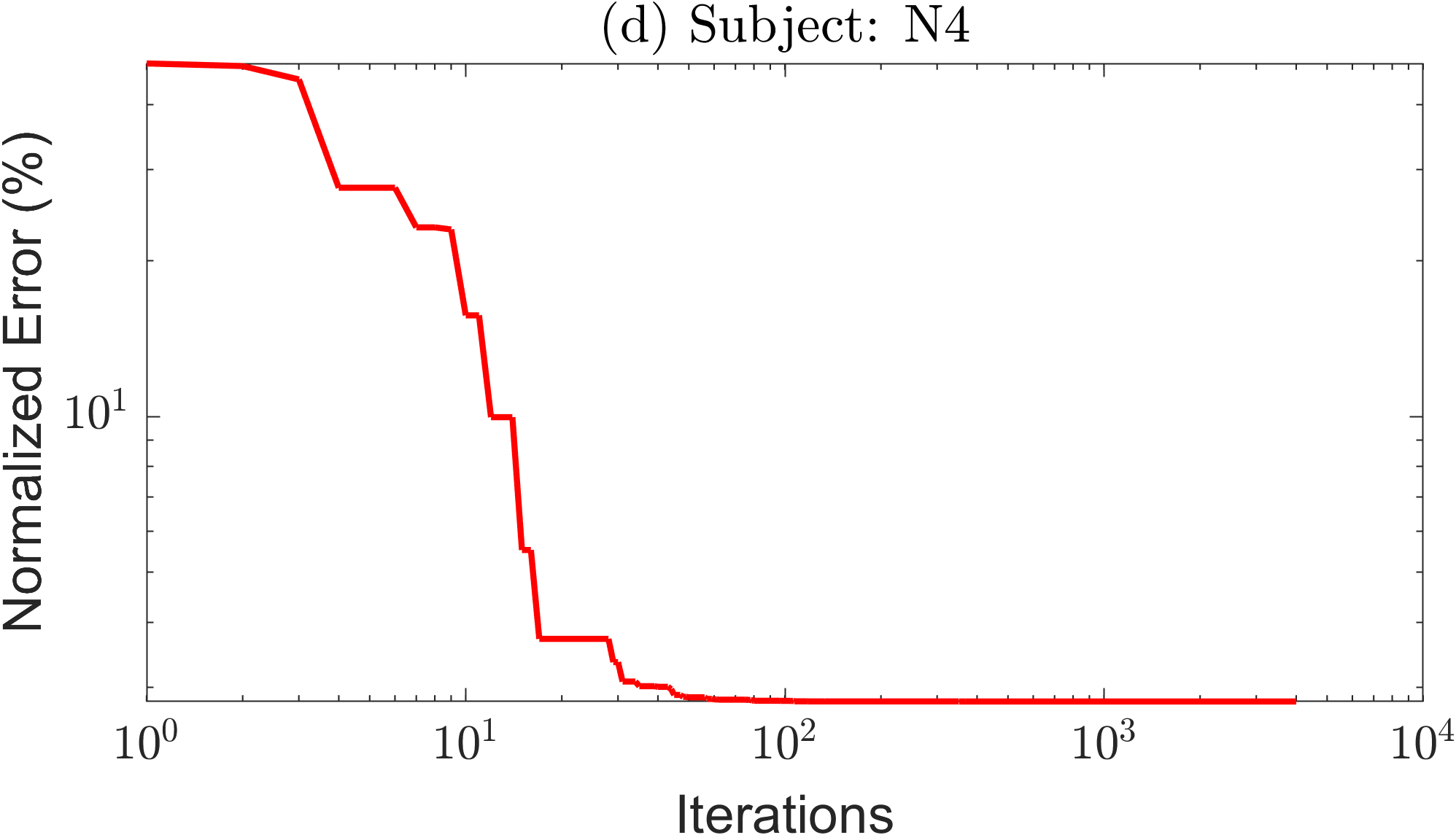}
    \caption{}
\end{subfigure}

\caption{Convergence plots of the optimization process for four subjects: (a) N1, (b) N2, (c) N3, and (d) N4, showing the normalized error between the experimental and calculated GAW over iterations. Both x and y axes are plotted on a logarithmic scale.}
\label{r1}
\end{figure}

During the optimization, the algorithm iteratively adjusts values of $S$, $K$, and $C_f$  to minimize the normalized error (referring to Eq. \ref{eq23}) between the calculated and experimental GAW. Fig. \ref{r1} illustrates that the normalized error decreases rapidly in the initial iterations. After this part, the rate of reduction slows significantly, and beyond a certain number of iterations, the error stabilizes. This plateau at the end indicates that the PSO algorithm has converged. For all subjects, the solution converges after approximately 1,000 iterations. 

\begin{table}[htpb]
\centering
\caption{Optimized parameter values and corresponding normalized errors for four subjects.}
\begin{tabular}{c c c c c}  \label{T4}
\\ \hline
Subject & $S$($\frac{\text{pixels}}{\text{mm}^2}$) & $K$ ($\frac{N}{m}$) & $C_f$ & Normalized Error (\%)\\ \hline
N1 & 43.7988 & 396.6550 & 7536.2 &1.8685\\
N2 & 70.7391 & 113.9710 & 11250 &1.9959\\
N3 & 24.9649 & 671.1919 & 489.5168 &2.1039\\
N4 & 29.6812 & 216.0464 & 829.8627 &2.8136\\ \hline
\end{tabular}

\end{table}

The optimal parameter values for different subjects are obtained at the end of the optimization process as shown in Table \ref{T4}. The scaling factor ranges from 24.9 to 70.7, $K$ from 113.97 to 671.19 $\frac{N}{m}$, and $C_f$ from 489.5 to 11,250. Subject N2 has the highest scaling factor and $C_f$ but the lowest $K$, while subject N3 has the lowest scaling factor and $C_f$ but the highest $K$. The normalized error ranges from 1.87\% to 2.8136\%, with subject N4 having the highest error. The minimal normalized error values ($<$3\%) indicate that the model provides a good approximation of the experimental results. 
\begin{figure}[htbp]
\centering

\begin{subfigure}{0.45\textwidth}
    \centering
    \includegraphics[width=\linewidth]{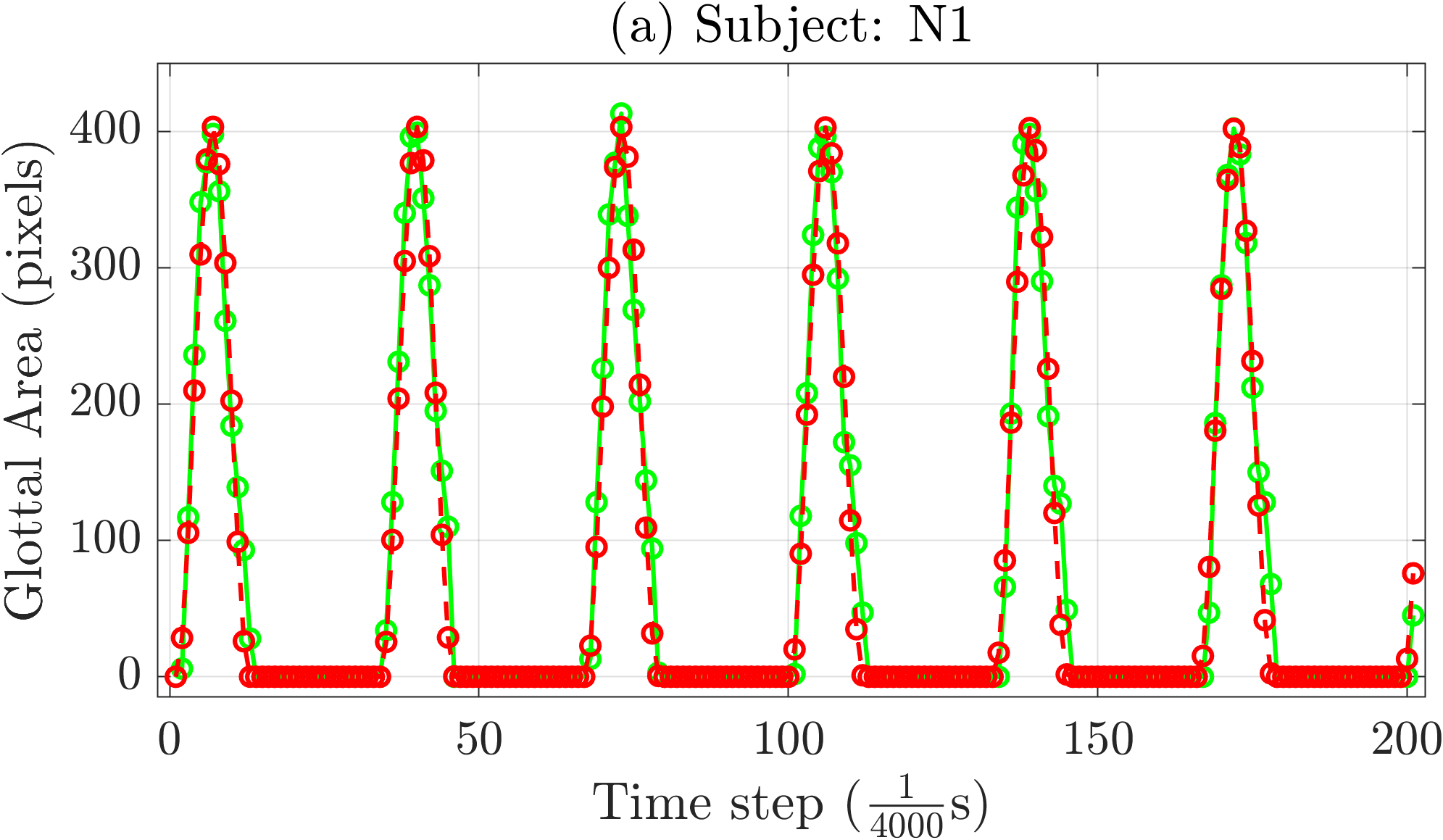}
    \caption{}
\end{subfigure}
\hfill
\begin{subfigure}{0.45\textwidth}
    \centering
    \includegraphics[width=\linewidth]{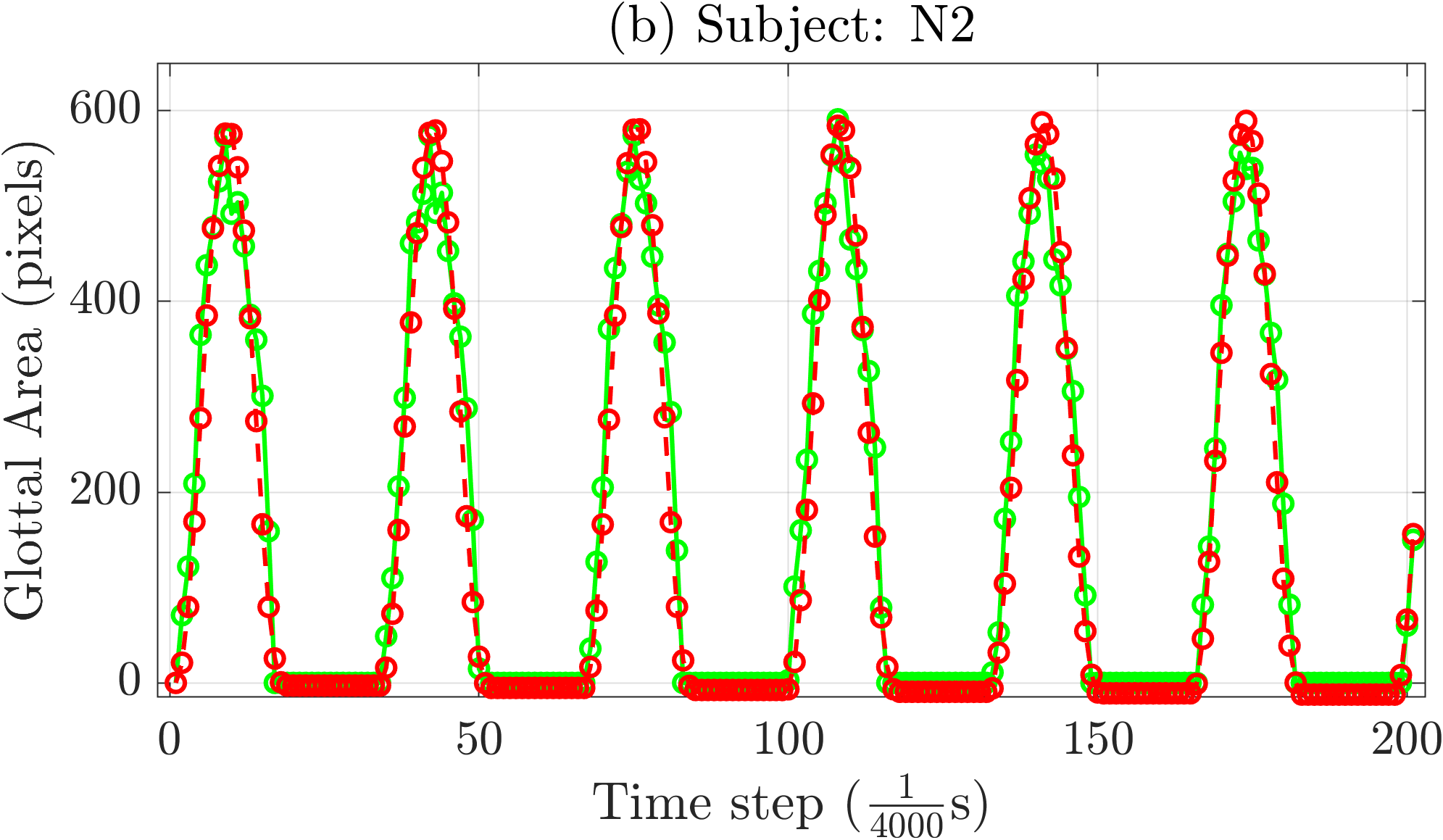}
    \caption{}
\end{subfigure}

\vspace{0.5cm}

\begin{subfigure}{0.45\textwidth}
    \centering
    \includegraphics[width=\linewidth]{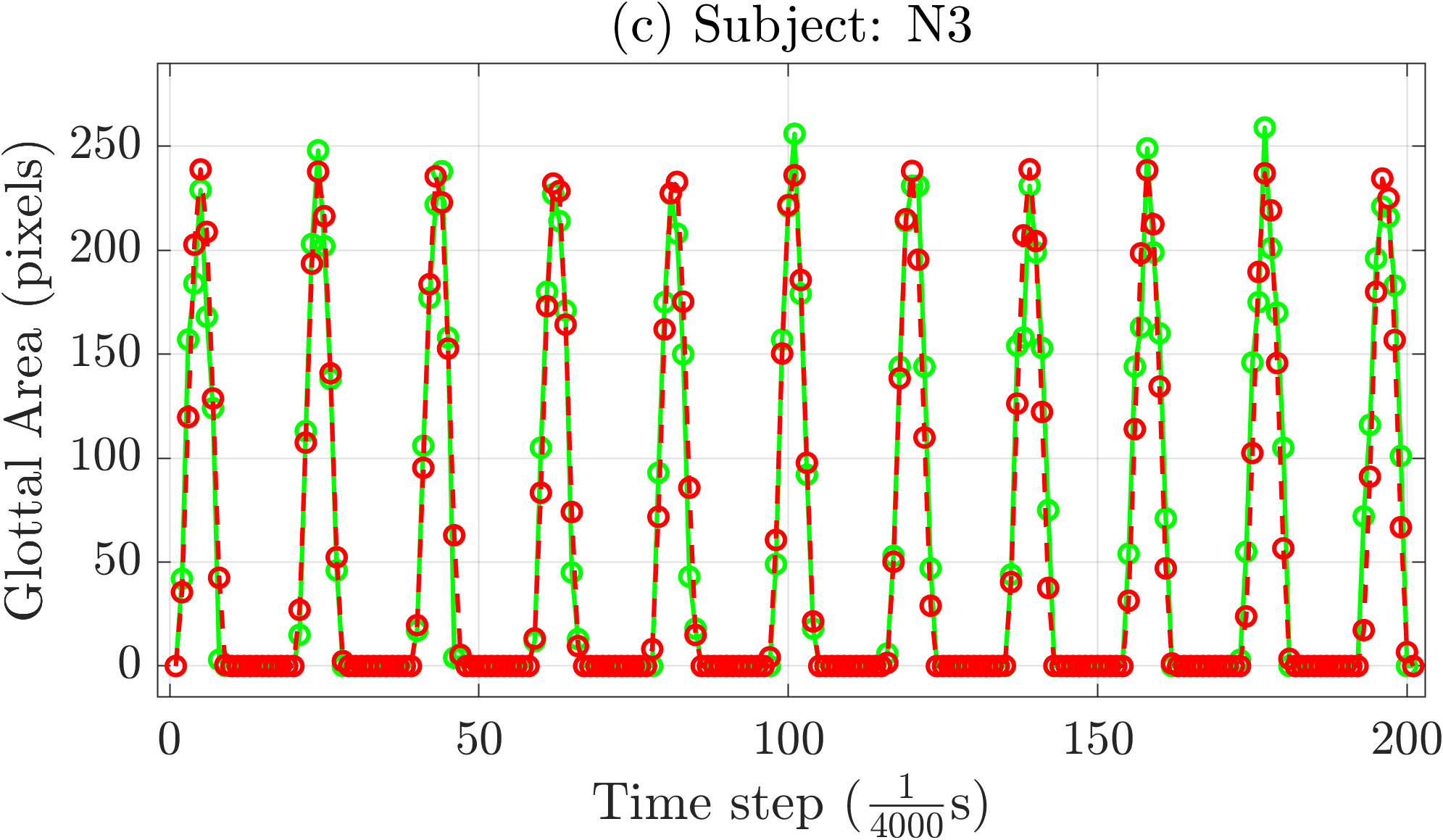}
    \caption{}
\end{subfigure}
\hfill
\begin{subfigure}{0.45\textwidth}
    \centering
    \includegraphics[width=\linewidth]{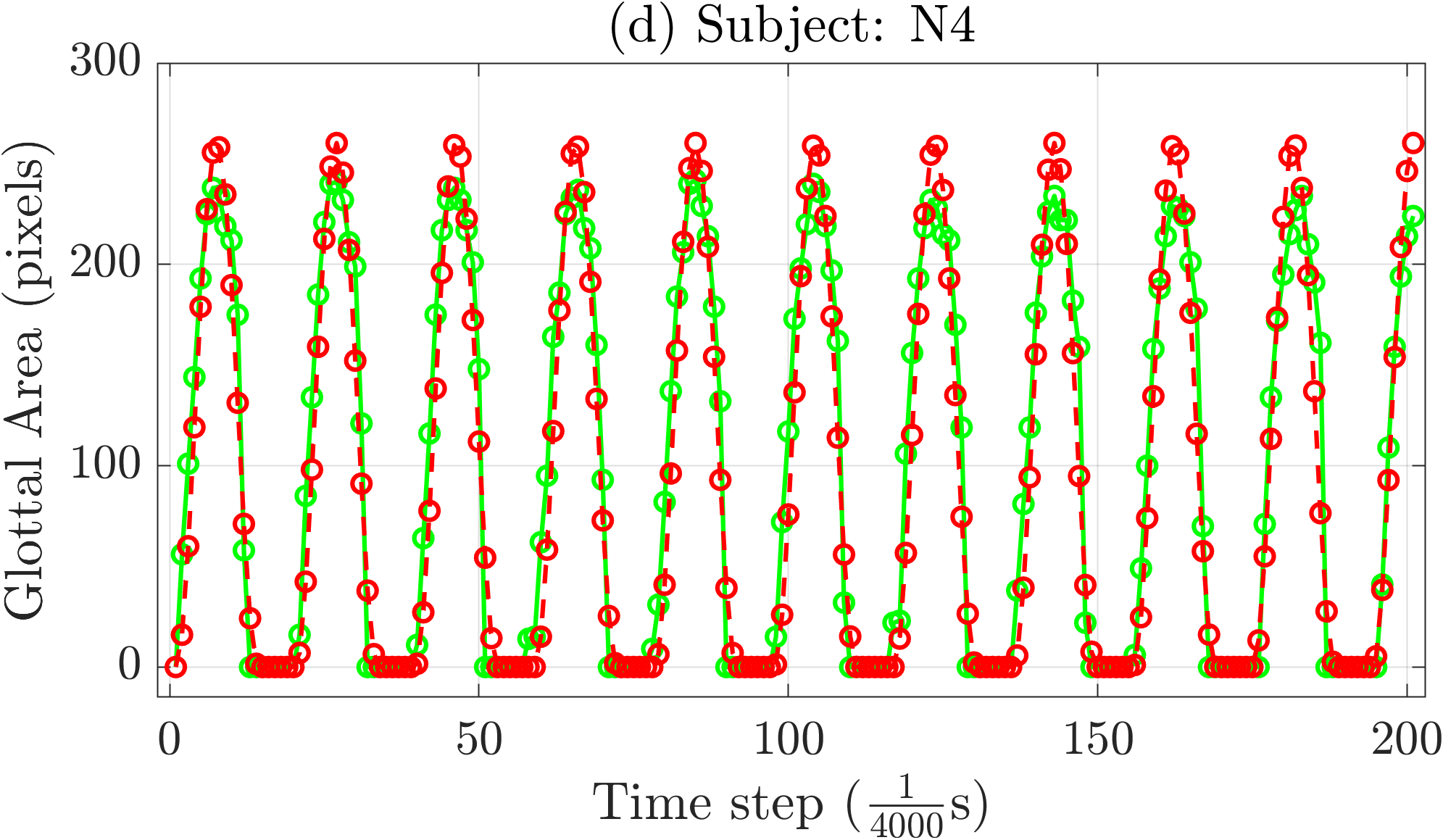}
    \caption{}
\end{subfigure}

\caption{Comparison of GAW from experimental data (green) and numerical simulations with optimized parameters (red) for different subjects. The plots correspond to subject N1 (upper left), N2 (upper right), N3 (lower left), and N4 (lower right). }
\label{r2}
\end{figure}

Using the optimized parameters, the GAW for each subject is calculated and compared with the corresponding experimental data, as shown in Fig. \ref{r2}. It is apparent that the calculated GAWs sometimes underestimate or overestimate the experimental data. This occurs because the simulated GAW is periodic, whereas the experimental data is quasi-periodic, with slight variations in the amplitude, closure time, and cycle length within different cycles. However, the calculated data closely follows the experimental data with minor discrepancies. This agreement confirms that the computational approach accurately models the vocal folds vibrations.
\begin{figure}[htbp]
\centering

\begin{subfigure}{0.45\textwidth}
    \centering
    \includegraphics[width=\linewidth]{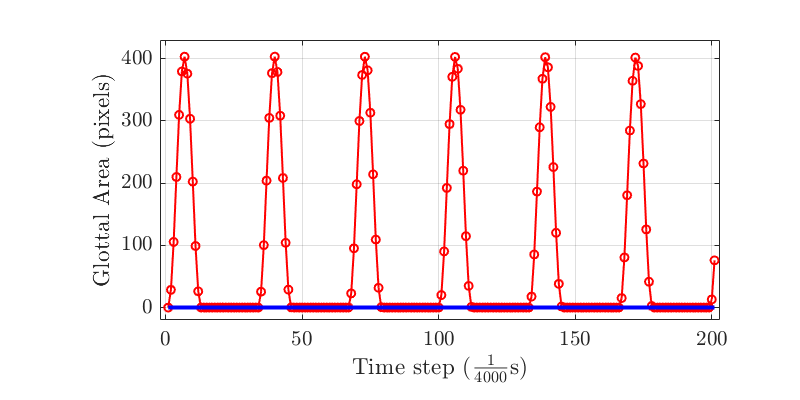}
    \caption{}
\end{subfigure}
\hfill
\begin{subfigure}{0.45\textwidth}
    \centering
    \includegraphics[width=\linewidth]{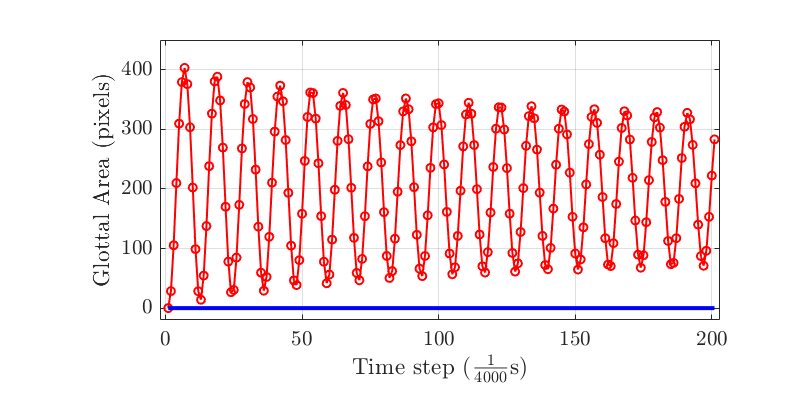}
    \caption{}
\end{subfigure}
\caption{GAW for subject N1 calculated with optimized parameters. The left plot represents the model with the flow separation, while the right plot represents the model without the flow separation. The blue line indicates the zero glottal area. }
\label{r3}
\end{figure}

As previously discussed in the Model Parameters section, the damping coefficient is zero during the open phase but it becomes non-zero in the closing, closed, and opening phases. The vocal folds must overcome the damping in these later phases to achieve sustained oscillation. The energy loss due to damping is recovered through the force asymmetry generated by the flow separation force during the closing phase. Without the flow separation, there is no mechanism to compensate for the lost energy. In this case, the vocal folds cannot fully overcome the damping force during the closing phase. As a result, they begin to open again before reaching $x_0$, preventing the closure. Due to the inability to overcome energy loss, the vocal folds continuously lose energy during the closing and opening phases of the cycle, causing the oscillation amplitude to decrease with each cycle, as shown in Fig. \ref{r3}.

Even with the flow separation included in the model, some amount of energy can still be dissipated through damping. If the vocal folds do not complete the closing phase, meaning they fail to reach $x_0$ at the end of the closing phase and start separating before the closure, they cannot fully recover the energy loss to the damping. The vocal folds must complete the closing phase and transition into the closed phase to completely regain the energy loss.

The energy loss due to damping exactly equals the energy gained through force asymmetry if the glottal area remains zero during the closed phase, meaning the vocal fold velocity is zero during the transition from the closing to closed phase. As a result, the vocal folds neither lose nor gain energy in each cycle, keeping the oscillation amplitude constant. Conversely, extra energy is added in each cycle, if the glottal area is negative during the closed phase due to the vocal folds having nonzero velocity while shifting from the closing to closed phase. In this case, the oscillation amplitude increases with each cycle. In the simulation, the optimal $C_f$ and $K$ values correspond to a zero $A_g$ value throughout the entire closure.

\begin{figure}[htbp]
\centering

\begin{subfigure}{0.45\textwidth}
    \centering
    \includegraphics[width=\linewidth]{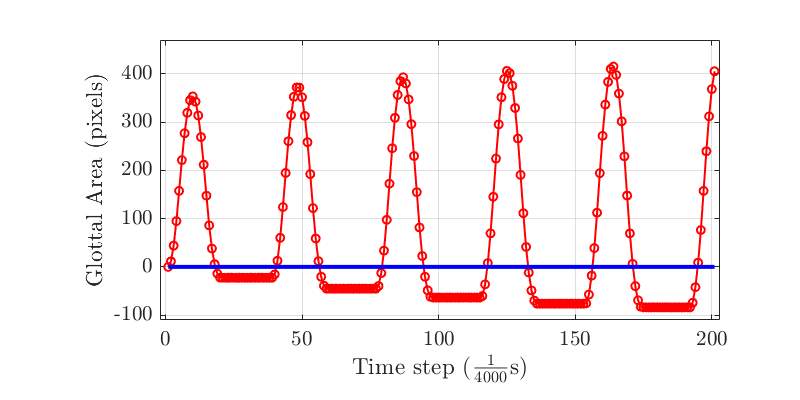}
    \caption{}
\end{subfigure}
\hfill
\begin{subfigure}{0.45\textwidth}
    \centering
    \includegraphics[width=\linewidth]{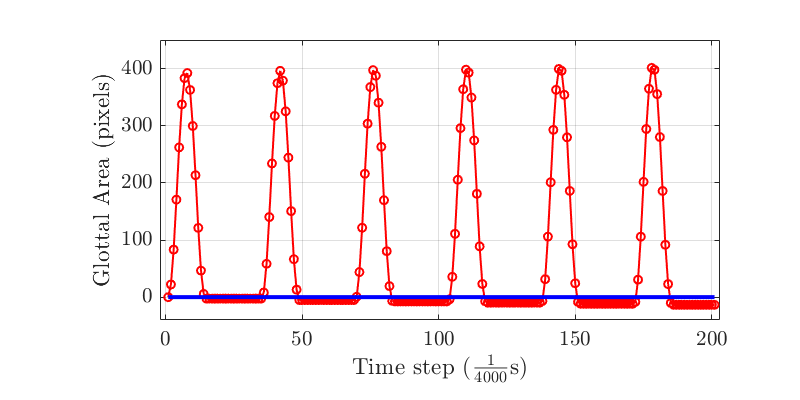}
    \caption{}
\end{subfigure}

\vspace{0.5cm}

\begin{subfigure}{0.45\textwidth}
    \centering
    \includegraphics[width=\linewidth]{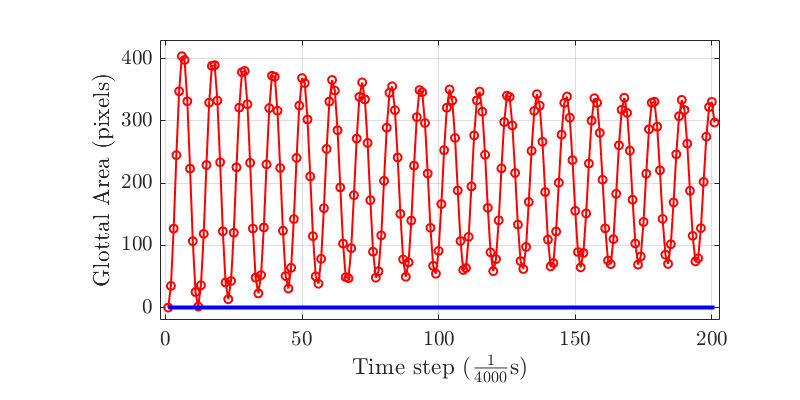}
    \caption{}
\end{subfigure}
\hfill
\begin{subfigure}{0.45\textwidth}
    \centering
    \includegraphics[width=\linewidth]{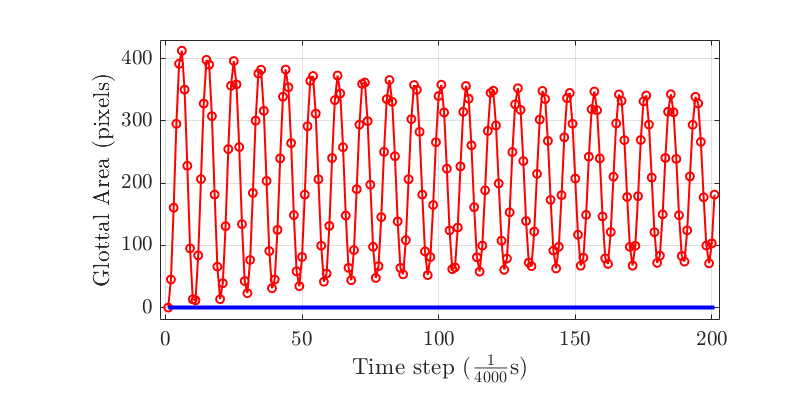}
    \caption{}
\end{subfigure}

\caption{GAW for subject N1 calculated with 80\% decrease (upper left), 30\% decrease (upper right), 30\% increase (lower left), and 80\% increase (lower right) in $K$ relative to its optimal value. All other parameters are set to their optimal values. The blue line indicates the zero glottal area. }
\label{r4}
\end{figure}

Changing $K$ significantly affects the frequency of the glottal cycles, as shown in Fig. \ref{r4}. $K$ influences the duration of all phases except the closed phase, where the elastic force is canceled due to the inclusion of the structural contact force (previously discussed in the modeling framework section). In general, as $K$ increases, the number of cycles per unit time also increases. If the optimal $C_f$ is used, using a $K$ value higher than the optimal $K$ will prevent the closure. In this case, excessive elastic force in the closing phase stops the vocal folds from reaching $x_0$. As a result, the oscillation amplitude will decrease over time, and closure will not occur. On the other hand, if $K$ is decreased below its optimal value, the vocal fold velocity will be greater than zero when moving from the closing to the closed phase. Consequently, the oscillation amplitude will increase over time.

\begin{figure}[htbp]
\centering

\begin{subfigure}{0.45\textwidth}
    \centering
    \includegraphics[width=\linewidth]{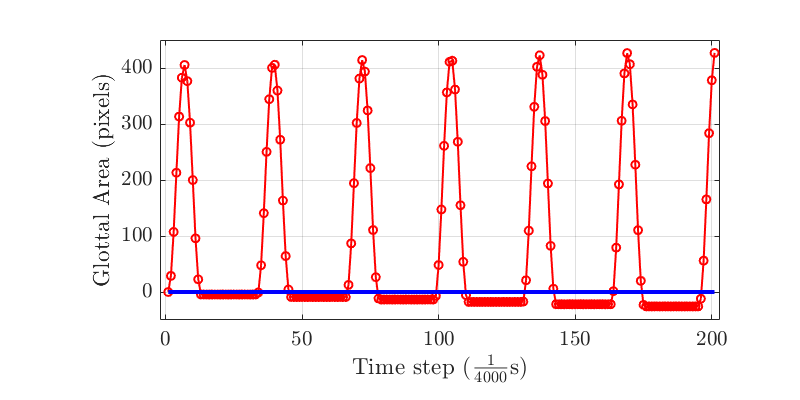}
    \caption{}
\end{subfigure}
\hfill
\begin{subfigure}{0.45\textwidth}
    \centering
    \includegraphics[width=\linewidth]{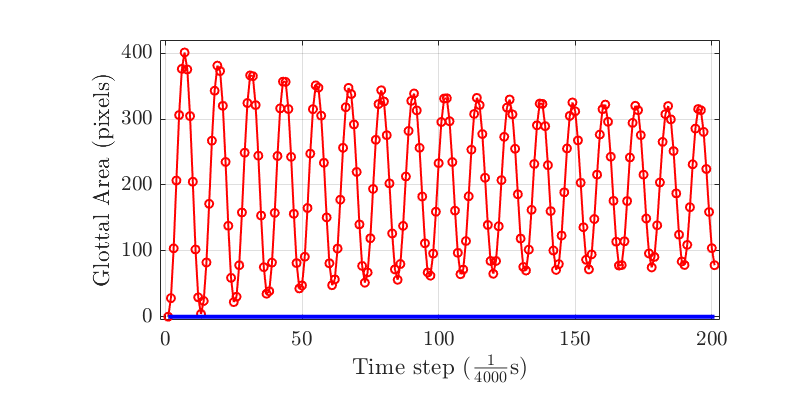}
    \caption{}
\end{subfigure}

\caption{GAW for subject N1 calculated with 30\% decrease (left) and 30\% increase (right) in $C_f$ relative to its optimal value. All other parameters are set to their optimal values. The blue line indicates the zero glottal area.}
\label{r5}
\end{figure}

The damping has a considerable impact on the dynamics of the vocal folds, as illustrated in Fig. \ref{r5}. In this simulation, if $C_f$ is lower than the optimal value (Fig. \ref{r5}-a), the vocal folds will enter the closed phase with nonzero velocity, causing the oscillation amplitude to grow. Conversely, if $C_f$ is increased beyond its optimal value (Fig. \ref{r5}-b), the vocal folds cannot transition from the closing phase to the closed phase, resulting in a damped oscillation without closure.  

\begin{figure}[htbp]
\centering

\begin{subfigure}{0.45\textwidth}
    \centering
    \includegraphics[width=\linewidth]{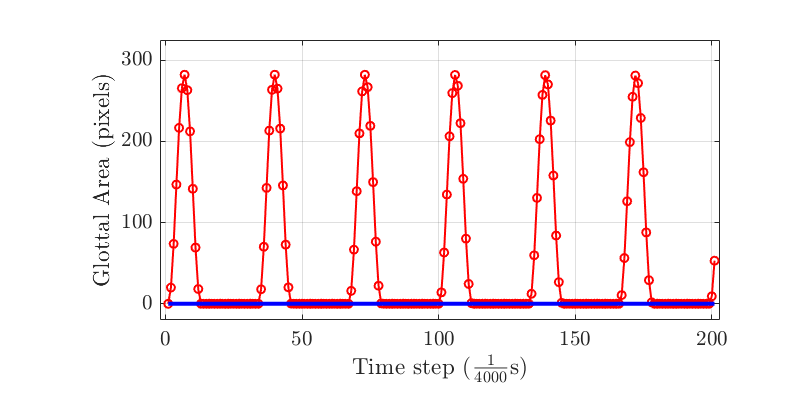}
    \caption{}
\end{subfigure}
\hfill
\begin{subfigure}{0.45\textwidth}
    \centering
    \includegraphics[width=\linewidth]{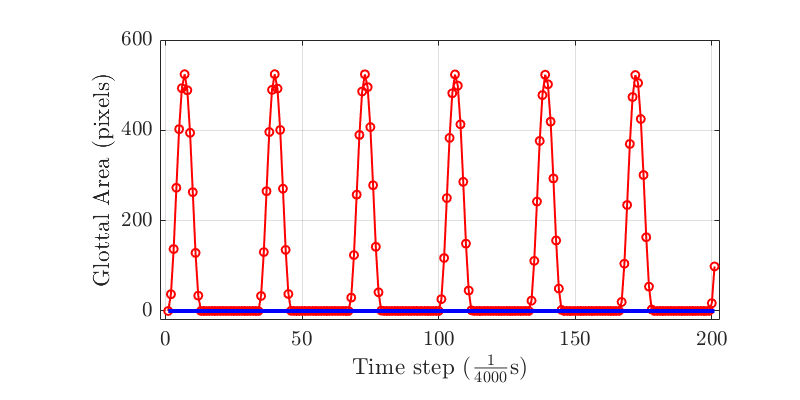}
    \caption{}
\end{subfigure}

\caption{GAW for subject N1 calculated with 30\% decrease (left) and 30\% increase (right) in the scaling factor relative to its optimal value. All other parameters are set to their optimal values. The blue line indicates the zero glottal area.}
\label{r6}
\end{figure}

Changing the scaling factor does not alter the GAW shape or the time period of any phase (see Fig. \ref{r6}). It only scales the displacements up or down linearly based on the scaling factor.

\newpage
\newpage
\newpage

\section{Discussion}
This study presents a simplified yet physically consistent framework for modeling vocal fold dynamics, demonstrating that a single-degree-of-freedom system can reproduce experimentally observed glottal area waveforms and capture key features of sustained phonation.\\ 
The damping amplification factor plays a critical role in regulating vocal fold closure and, consequently, the overall oscillatory dynamics of the proposed model. Because the current single-degree-of-freedom framework cannot explicitly model tissue deformation and overlap during collision, unlike higher-order models based on Hertzian contact mechanics (\citeauthor{i25}, \citeyear{i25}; \citeauthor{i26}, \citeyear{i26}), the damping amplification factor was optimized to achieve physiologically realistic closure with minimal vocal fold overlap. Since the degree of overlap cannot be determined directly from the experimental glottal area waveforms, the optimization targeted a glottal area that remained as close to zero as possible throughout the closed phase while avoiding excessive compression.
A damping amplification factor greater than its optimized value causes the vocal folds to decelerate to zero velocity before reaching the full-closure position ($x_0$). Because the net force at this point is positive, the vocal folds accelerate in the opening direction and begin to separate before achieving complete closure. Conversely, a damping amplification factor smaller than the optimized value causes the vocal folds to reach zero velocity only after moving beyond $x_0$, resulting in excessive compression and overlap along the midline, which is reflected as a negative glottal area during the closed phase. When the damping amplification factor is equal to its optimized value, the vocal folds reach zero velocity at a position very close to $x_0$, thereby achieving complete closure with only a negligible negative glottal area and minimal overlap.

The inclusion of moving separation point during the closing phase was crucial for achieving vocal fold closure and maintaining sustained phonation. Without this mechanism, even a small amount of damping causes energy loss across cycles, preventing the vocal folds from reaching the closed phase. 
During the open phase, no energy is dissipated because the damping force is assumed to be zero. In the initial part of the closed phase, some additional energy is dissipated due to the presence of the damping force. However, this energy loss does not prevent the vocal folds from achieving full closure; instead, it only influences the degree of compression during the early closed phase. In the absence of this additional energy dissipation, the vocal folds would still achieve complete closure but with no compression and, consequently, no negative glottal area.

During the later part of the closing phase, the total force becomes zero due to the structural reaction force during the collision and the zero velocity of the vocal folds. Thus, the overall effect of passing through the closed phase does not contribute to any loss of force required for achieving closure in the next cycle. So, the only mechanism of loosing energy in each cycle is the damping force acting during the closing and opening phases of the oscillation system. During the closing phase, the aerodynamic force acts opposite to the direction of vocal fold displacement, resulting in negative work, whereas during the opening phase, the aerodynamic force and vocal fold displacement are aligned, producing positive work. This follows from the assumption that $F_{s.a.}$ such that the transition from the opening to the closing phase occurs when the aerodynamic force changes sign and passes through zero. If the positive and negative work are equal, the vocal folds will gradually lose energy because no additional positive work is available to compensate for the energy dissipated by damping. Consequently, the displacement of the vocal folds during both the opening and closing phases would decrease with each oscillation cycle. To compensate for this energy loss, an asymmetry in the aerodynamic forces during the opening and closing phases is required, such that the positive work exceeds the negative work. The flow separation during the closing phase can address this issue by reducing the negative work, thereby generating an additional positive balancing force that can counteract the energy loss due to damping.

Differences between the numerical and experimental results can be attributed to several factors. The use of linear spring elasticity represents a simplification, as vocal fold tissues exhibit highly nonlinear elastic behavior, leading to some discreipancies. Similarly, modeling changes in the damping coefficient only during the closing, closed, and opening phases may not fully capture the nonlinear damping behavior of the vocal folds during the open phase, which constitutes a substantial portion of the glottal cycle. As a first step, the proposed model assumed periodic and symmetric vocal fold dynamics, with the dynamic parameters repeating the same pattern from one glottal cycle to the next. Future work will extend the model to incorporate cycle-to-cycle variability and asymmetric vocal fold behavior, enabling the study of irregular and asymmetric phonation patterns. Our work provides a foundation for developing an integrated lumped-element model that includes all voice production subsystems and generates subject-specific inlet conditions for computational models of phonation (\citeauthor{i33}, \citeyear{i33}).

\section{Conclusions}
This study proposed a lumped model with a single mass-spring-damper to properly capture the physics of vocal folds vibration. The proposed model addressed the shortcomings of the single degree of freedom models by incorporating a flow separation and structural force in its construction. The model incorporated a moving separation point during the closing phase to account for the progressive reduction in the effective length of the vocal folds over which the aerodynamic force acts. With this approach, the sustained vibration of the vocal folds with full closure during the closed phase was achieved without considering the source-tract interactions. In addition, the damping coefficient was modeled as a displacement-dependent parameter that increased during the closing and closed phases to represent the progressive increase in mechanical resistance as the vocal folds approached contact and underwent compression. An external structural force was incorporated during the vocal folds closed phase to maintain closure for a prescribed period while accounting for the reaction force generated by the collision of vocal folds. Thus, despite having only a single degree of freedom, the incorporation of these mechanisms enables a more physiologically realistic representation of vocal fold dynamics while improving overall model accuracy. The model accurately reproduced the experimentally observed glottal area waveforms in HSV data across subjects, with normalized errors below 3\%. Owing to its simplicity and computational efficiency, the proposed model provides a practical framework for investigating the biomechanical and aerodynamic mechanisms underlying sustained vowel phonation. By requiring only a single degree of freedom and a minimal set of parameters, the model facilitates the study of the fundamental forces governing vocal fold vibration while reducing computational complexity.
One prominent approach to further extend the proposed modeling paradigm is to consider the power-law rheology and viscoelasticity of bio-tissues. This will be looked into in our future works, following our earlier developmental research in applied and computational mathematics (\citeauthor{i28}, \citeyear{i28}; \citeauthor{i29}, \citeyear{i29}; \citeauthor{i30}, \citeyear{i30}; \citeauthor{i31}, \citeyear{i31}; \citeauthor{i32}, \citeyear{i32}).

\section*{Acknowledgements}
The authors would like to acknowledge the support from the National Institutes of Health (NIH) National Institute on Deafness and Other Communication Disorders (NIDCD) under awards R21DC020003 and K01DC017751, the U.S. Army Research Office (ARO) Young Investigator Program (YIP) under award W911NF-19-1-0444, the National Science Foundation (NSF) under award DMS-1923201, and Michigan State University Discretionary Funding Initiative. In addition, the authors would like to thank Drs. Dimitar D. Deliyski and Stephanie R.C. Zacharias for their help and support with the data collection, and the Institute for Cyber-Enabled Research (ICER) at Michigan State University for providing computational resources and facilities.

\bibliography{citation}

\end{document}